\documentclass[aps,prd,twocolumn,showpacs,groupedaddress,nofootinbib,amsmath]{revtex4}
\usepackage{graphicx}
\usepackage{epsfig}
\usepackage{multirow}
\usepackage{subfigure}
\newcommand{\be}{\begin{equation}}
	\newcommand{\ee}{\end{equation}}
\newcommand{\ba}{\begin{eqnarray}}
	\newcommand{\ea}{\end{eqnarray}}

\newcommand{\Dslash}{\lower-0.18ex\hbox{\makebox[-1pt][l]{\,/}}D}

\begin{document}
	
	\title{Masses of the conjectured H-dibaryon at different temperatures  }

	\author{Liang-Kai Wu}
	\affiliation{School of Physics and Electronic Engineering, Jiangsu University, Zhenjiang, 212013, People's Republic of China}
	
	\author{Han Tang}
	\thanks{Corresponding author. Email address: 18570503013@163.com}
	\affiliation{School of Physics and Electronic Engineering, Jiangsu University, Zhenjiang, 212013, People's Republic of China}
	
	\author{Xin-Yang Wang}
	\affiliation{Center for Fundamental Physics, School of Mechanics and Physics, Anhui University of Science and Technology, Huainan, Anhui 232001, People's Republic of China}
	
	\author{Ning Li}
	\thanks{Corresponding author. Email address: lining@ujs.edu.cn}
	\affiliation{School of Physics and Electronic Engineering, Jiangsu University, Zhenjiang, 212013, People's Republic of China}
	
	\date{\today}
	
	\begin{abstract} 
We present a lattice QCD determination of masses of the conjectured H-dibaryon $m_H$ at nine different temperatures $T/T_c =0.24, 0.63, 0.76, 0.84, 0.95, 1.09, 1.27, 1.52, 1.90$.
In the meantime,  the masses of baryon $N$,  $\Sigma$, $\Xi$ and $\Lambda$ at different temperatures
		are also computed.
		The simulation is performed on anisotropic lattice with $N_f=2+1$ flavours of clover fermion  at quark mass
		which corresponds to  $m_\pi=384(4) {\rm MeV} $.  The thermal ensembles were provided by the FASTSUM
		collaboration and the zero temperature ensembles by the Hadspec collaboration.  We also calculate  the spectral density of the correlation function of those particles.
		The spectral density distributions show rich peak structure at the lowest temperature, while at intermediate temperatures, the mass values of those particles obtained by extrapolation method reflect a two-peak structure.
		While the spectral density for octet baryon becomes smooth at $T/T_c = 1.27, 1.52, 1.90$,  the spectral density for H-dibaryon becoms smooth at $T/T_c = 1.90$.
		At $T/T_c =0.24 $, the mass difference of H-dibaryon
		and
		$\Lambda$ pair   $\Delta m = m_H - 2\,m_{\Lambda} $  is estimated to be $\Delta m = -14.6(6.2) {\rm MeV}$ which suggests there exists a  bound  H-dibaryon state.	
	\end{abstract}
	\maketitle

	\section{INTRODUCTION}
	\label{SectionIntro}
	
	Quantum chromodynamics (QCD) describes the dynamics of quarks and gluons. It underlies all of nuclear physics from  hadronic mass
	spectrum to the phase transition of hadronic matter to quark-gluon plasma (QGP).  Because of the nature of the strong interaction of QCD at low
	energy scale, the perturbative method cannot be applied to explain those low energy phenomena of nuclear physics. Fortunately,  lattice QCD
	which is based on  first principles can be employed to make precise predictions for hadronic quantities, especially, for those phenomena
	which are difficult to explore in  the laboratory, for example,  mass spectrum of baryon at different high temperatures.
	
	In 1976, by using the bag model, Jaffe predicted a flavour-singlet state ($uuddss$) with quantum number $I(J^P)=0(0^+) $ which is called
	H-dibaryon~\cite{Jaffe:1976yi}. In contrast with the only known stable dibaryon (deuteron) whose binding energy is about $2.2\ {\rm MeV}$, Jaffe predicted that the binding energy
	of H-dibaryon is about $80\ {\rm MeV}$ below the $\Lambda\Lambda$ threshold $2230\ {\rm MeV}$ which means that H-dibaryon is a deeply bound state.
	
	Unlike  mesons and baryons, this exotic hadron may be relevant to the hypernuclei, and to  the strange matter that could exist
	in the core of neutron stars.  Moreover, it is a potential candidate for dark matter~\cite{Farrar:2017eqq}. As a consequence,
	this prediction triggered a vigorous search for such a state, both experimentally~\cite{Aoki:1991ip,KEK-PSE224:1998trj,Takahashi:2001nm,Yoon:2007aq,KEKE176:2009jzw,Nakazawa:2010zza,Belle:2013sba,BaBar:2018hpv,Ekawa:2018oqt}
	and theoretically~\cite{Iwasaki:1987db,Luo:2011ar,Luo:2007zzb,Mackenzie:1985vv,Pochinsky:1998zi,Wetzorke:1999rt,Wetzorke:2002mx,Beane:2009py,NPLQCD:2010ocs,
		Beane:2011zpa,NPLQCD:2011naw, NPLQCD:2012mex,Inoue:2010hs,Inoue:2010es,Inoue:2011ai,HALQCD:2019wsz,Sasaki:2016gpc,HALQCD:2018lur,Francis:2018qch,Green:2021qol}.

	The observation of double hypernuclei, $^A_{\Lambda\Lambda}Z$, is very important in connection with the existence of
	H-dibaryon~\cite{Aoki:1991ip}. If the mass of H-dibaryon $m_H$ is much smaller than the mass of double $\Lambda$ hyperon $2\,m_{\Lambda}$,
	a double hypernuclei may decay into H-dibaryon and a residual nucleus by strong interation.
	In such case, the branching  ratio for the decay of double  hypernuclei through weak interaction is very small,
	practically, cannot be observed~\cite{Aoki:1991ip}. So the observation of the weak decay of  double  hypernuclei will
	put limitation on the mass of H-dibaryon $m_H > 2\,m_{\Lambda} - B_{\Lambda\Lambda}$ with $B_{\Lambda\Lambda}$ being
	the binding energy of $\Lambda\Lambda$ hyperons.
	
	The experiments~\cite{Aoki:1991ip,KEK-PSE224:1998trj,Takahashi:2001nm,Yoon:2007aq,KEKE176:2009jzw,Nakazawa:2010zza,Ekawa:2018oqt}
	investigated the nuclear capture of $\Xi^-$ at rest produced in $(K^-,K^+)$ reaction, and
	observed the sequential weak decay of double  hypernuclei, then measured the binding energy and
	interaction energy of $\Lambda\Lambda$~\cite{Aoki:1991ip,Takahashi:2001nm,Nakazawa:2010zza,Ekawa:2018oqt}, or
	the cross section of enhanced production of $\Lambda\Lambda$ pair~\cite{KEK-PSE224:1998trj,Yoon:2007aq,KEKE176:2009jzw}.
	The results do not confirm the existence of H-dibaryon, and set the lower limit for the mass of H-dibaryon.
	
	The experiments~\cite{Belle:2013sba,BaBar:2018hpv} were carried out to search for H-dibaryon or deeply bound singlet $uuddss$ sexaquark  $S$
	(for the explanation of $S$,
	see~\cite{Farrar:2017eqq})
	in
	$\Upsilon \rightarrow S\bar\Lambda\bar\Lambda$ decay. Their results show no signal for the existence of H-dibaryon
	or $S$ particle.
	
	As a theoretically tool, Lattice QCD is also used to investigate H-dibaryon. Some quenched studies show that $m_H < 2\ m_{\Lambda}$,
	such results support that H-dibaryon is a bound state~\cite{Iwasaki:1987db,Luo:2011ar,Luo:2007zzb}
	, while other quenched studies show that H-dibaryon is not a bound state~\cite{Mackenzie:1985vv,Pochinsky:1998zi,Wetzorke:1999rt,Wetzorke:2002mx}.

	Aside from quenched studies, simulations with dynamical fermions have been carried out by NPLQCD, HALQCD and other groups. The NPLQCD collaboration
	investigated baryon-baryon scattering, and then, extracted the phase shift by employing L{\"u}scher's method~\cite{Luscher:1986pf,Luscher:1990ux} to
	distinguish scattering states from binding states~\cite{Beane:2003da,Beane:2006mx,Beane:2009py,NPLQCD:2010ocs,Beane:2011zpa,NPLQCD:2011naw,NPLQCD:2012mex}.
	They used this scenario
	to determine if there exists H-dibaryon, by making  simulation with $N_f=2+1$ dyamical fermions on anisotropic
	ensembles~\cite{Beane:2011zpa,NPLQCD:2010ocs,NPLQCD:2011naw}, and with $N_f=3$ dynamical fermions on isotropic ensembles~\cite{NPLQCD:2012mex}.
	
	The HALQCD collaboration investigated the baryon-baryon interaction in terms of the baryon-baryon potential. They extracted the Nambu-Bethe-Salpeter
	wave-function by computing the four-point green function on lattice, and then determined the baryon-baryon potential from the Nambu-Bethe-Salpeter
	wave-function. They used this method to address the existence of H-dibaryon on $N_f=3$ ensembles~\cite{Inoue:2010hs,Inoue:2010es,Inoue:2011ai},
	and on $N_f=2+1$ ensembles~\cite{HALQCD:2019wsz,Sasaki:2016gpc,HALQCD:2018lur}. Of the results obtained by the two groups, the simulations on
	$N_f=2+1$ ensembles~\cite{HALQCD:2019wsz,Sasaki:2016gpc,HALQCD:2018lur} by HALQCD claimed that H-dibaryon may be a $\Lambda\Lambda$ resonance.
	Other results by the two groups agreed on the presence of the H-dibaryon, despite disagreement on the binding energy~\cite{Francis:2018qch}.
	
	Ref.~\cite{Francis:2018qch} made simulations on ensembles of two dynamical quarks and one quenched strange quark. They applied L{\"u}scher's method to
	determine S-wave scattering phase shift with local and bilocal interpolators, and found  that for pion mass of 960 MeV, there exists a bound H-dibaryon.
	
	Ref.~\cite{Green:2021qol} carried out simulations by using O(a)-improved Wilson fermions at SU(3) symmetric point with $m_\pi=m_K\approx 420\,{ \rm MeV}$, and
	their results show that there exists a weakly bound H-dibaryon.
	
	Apart from the search for H-dibaryon, there are lattice QCD calculations for three-flavored heavy dibaryons~\cite{Junnarkar:2022yak,Mathur:2022ovu}.
	Three-flavored heavy dibaryons are states with possible quark flavour combinations with at least one of them as charm ($c$) or bottom ($b$) quark.
	
	Besides at zero temperature, the properties of hadrons at finite temperature are also one of the central goals of lattice QCD simulation
	(see, for example,~\cite{Aarts:2010ek,Aarts:2011sm,Aarts:2013kaa,Aarts:2014cda,Kelly:2018hsi,Aarts:2020vyb}). In the past decades, mesons
	at finite temperature have been studied
	extensively.  This is not the case for baryons. Baryons at finite temperature are hardly investigated on the lattice.
	In fact, there are a few lattice studies of baryonic screening and temporal masses \cite{DeTar:1987ar,Pushkina:2004wa,Aarts:2018glk,Datta:2012fz,Aarts:2015mma}. Nevertheless,
the behaviour of baryons in a hadronic medium
	is relevant to heavy-ion collisions. Therefore, there is a need to unambiguously understand the property  of baryons at finite temperature.

	Current research work
on H-dibaryon focuses on the problem of its existence at zero temperature from different aspects. In this paper, we make lattice QCD simulations to investigate the masses of the conjectured H-dibaryon and octet baryons at different temperatures. The change of mass of H-dibaryon with temperature is worth studying in its own right theoretically, moreover, the comparison of its mass
with $m_\Lambda$  can provide some information on its existence.
	
	The paper is organized as follows: In Sec.~\ref{SectionLattice},
	we present the technique details of the simulation which include the definition of correlation functions and the interpolating operators.
	Sec.~\ref{SectionSpectral} introduces the method of extracting spectral density from correlation function designed in Ref.~\cite{Hansen:2019idp}.
	Our simulation results are
	presented in Sec.~\ref{SectionMC},  followed by discussion in
	Sec.~\ref{SectionDiscussion}.

	\section{LATTICE CALCULATION AND SETUP}
	\label{SectionLattice}
	In our simulation, we compute the correlation functions of H-dibaryon and $\Lambda$, we also calculate the correlation function of  $N$, $\Sigma$ and $\Xi$.
	The generic form of correlation function is:
	\begin{eqnarray}
		\label{correlation_function}
		G(\vec{x},\tau) &= < O(\vec{x},\tau)O^+(0)>,
	\end{eqnarray}
	For the H-dibaryon interpolating operator, we choose the local operator. The starting point is the following operator notation for the different six quark combination~\cite{Francis:2018qch}:
	\begin{align}\label{eq:hexaquark}
		[abcdef] = \ & \epsilon_{ijk} \epsilon_{lmn} \Big( b^i C\gamma_5 P_+ c^j
		\Big)  \nonumber \\
		& \times\Big( e^l C\gamma_5 P_+ f^m \Big) \Big( a^k C\gamma_5 P_+ d^n \Big) ({\vec{x}}, t)\,,
	\end{align}
	where $a, b,\ldots,f$ denote generic quark flavors, and
	$P_+=(1+\gamma_0)/2$ projects the quark fields to positive parity.
	We choose the operator $O_\mathbf{H}$ as the H-dibaryon interpolating operator which transforms under the singlet irreducible representation of flavor SU(3)~\cite{Donoghue:1986zd,Golowich:1992zw,Wetzorke:1999rt,Wetzorke:2001tgi}:
	\begin{align}
		O_\mathbf{H} &= \frac{1}{48}\Big( [sudsud] - [udusds] - [dudsus]  \Big),
	\end{align}
	The H-dibaryon correlation function can be obtained based on the formulae in Ref.~\cite{Wetzorke:2001tgi}.
	
	For the baryon $\Lambda$ interpolating operator, we choose the standard definition
	which is given by (see, for example, Refs.~\cite{Leinweber:2004it,Montvay:1994cy,Gattringer:2010zz}):
	\begin{align}
		O_{\Lambda}(x)
		=& \frac{1}{\sqrt{6}} \epsilon_{abc}
		\left\{ 2
		\left( u^T_a(x)\ C \gamma_5\ d_b(x) \right) s_c(x)  \right.  \nonumber \\
		&   +\ \left( u^T_a(x)\ C \gamma_5\ s_b(x) \right) d_c(x)   \nonumber \\
		& \left. -\ \left( d^T_a(x)\ C \gamma_5\ s_b(x) \right) u_c(x)
		\right\}\ ,
		\label{chi1l8}
	\end{align}
	For the baryon $\Lambda$ correlation function, we choose the standard definition (see, for example, Refs.~\cite{Leinweber:2004it,Montvay:1994cy,Gattringer:2010zz}).

	For the baryon $N$, $\Sigma$  and $\Xi$, we take the standard definition of interpolating operator, and the corresponding definition of correlator (also, see, Refs.~\cite{Leinweber:2004it,Montvay:1994cy,Gattringer:2010zz}).
	
	After we get the correlation function, the mass can be obtained by fitting the exponential ansatz:
	\begin{align}
		\label{eq:Ansatz}
		G(\tau) =  A_{+} e^{-m_{+}\tau} +  A_{-} e^{-m_{-}(1/T-\tau)},
	\end{align}
	with  $m_{+}$ being the mass of the particle of interest, and $\tau$ in the interval $0 \leq \tau<1/T$ on a lattice at finite temperature $T$.

\begin{table*}[t]
	\caption{Parameters in the lattice action. This table is recompiled from Ref.~\cite{Aarts:2020vyb}.}
	\label{tab:parameters}
	\begin{ruledtabular}
		\centering
		\begin{tabular}{ l  l }
			gauge coupling (fixed-scale approach)     & $\beta = 1.5$ \\
			tree-level  coefficients 					& $c_0=5/3,\,c_1=-1/12$ \\
			bare gauge, fermion anisotropy 			& $\gamma_g = 4.3$, $\gamma_f = 3.399$ \\
			ratio of bare anisotropies          &  $\nu = \gamma_g / \gamma_f = 1.265$ \\
			spatial tadpole (without, with smeared links) & $u_s = 0.733566$, $\tilde{u}_s = 0.92674$ \;  \\
			temporal tadpole (without, with smeared links) \; & $u_\tau = 1$, $\tilde u_\tau = 1$  \\
			spatial, temporal clover coefficient 			& $c_s = 1.5893$, $c_\tau = 0.90278$ \\
			stout smearing for spatial links 			& $\rho = 0.14$, isotropic, 2 steps \\
			bare light quark mass for Gen2 		& $\hat m_{0, \rm light} = -0.0840$ \\
			bare strange quark mass          		& $\hat m_{0, \rm strange} = -0.0743$ \\
			light quark hopping parameter for Gen2 &  $\kappa_{\rm light} = 0.2780$ \\
			strange quark hopping parameter              &  $\kappa_{\rm strange} = 0.2765$ \\
		\end{tabular}
	\end{ruledtabular}
\end{table*}
	
	 In order to get the ground state energy of the particle concerned, it is best to choose large time extent lattice.  However, at finite temperature, if we make simulation on large time extent lattice, the lattice spacing
  must be chosen to be small.
  Therefore, it is  a dilemma for us to make lattice simulation at finite temperature presently.
    In order to get the ground state energy  as possible as we can on a relatively small time extent lattice,
	it is expedient for us to take an extrapolation method in our procedure to get the ground state mass.  We
	fit equation~(\ref{eq:Ansatz}) to correlators in a series of time range $[\tau_1,\tau_2]$ where $\tau_2$ is fixed to the  whole time extent,  and $\tau_1$ runs over several values from  $\tau_1=1, 2, 3, 4,...$, then we can get a series of mass values
	which correspond to different early Euclidean time slices suppression.
After that,
	we plot the mass values obtained in different time interval $[\tau_1,\tau_2]$  against $1/\tau_1$,  and fit a linear expression to those mass values, then, extrapolate the linear expression  to $\tau_1 \to \infty$.

	\begin{figure*}[t!]
		\includegraphics*[width=0.49\textwidth]{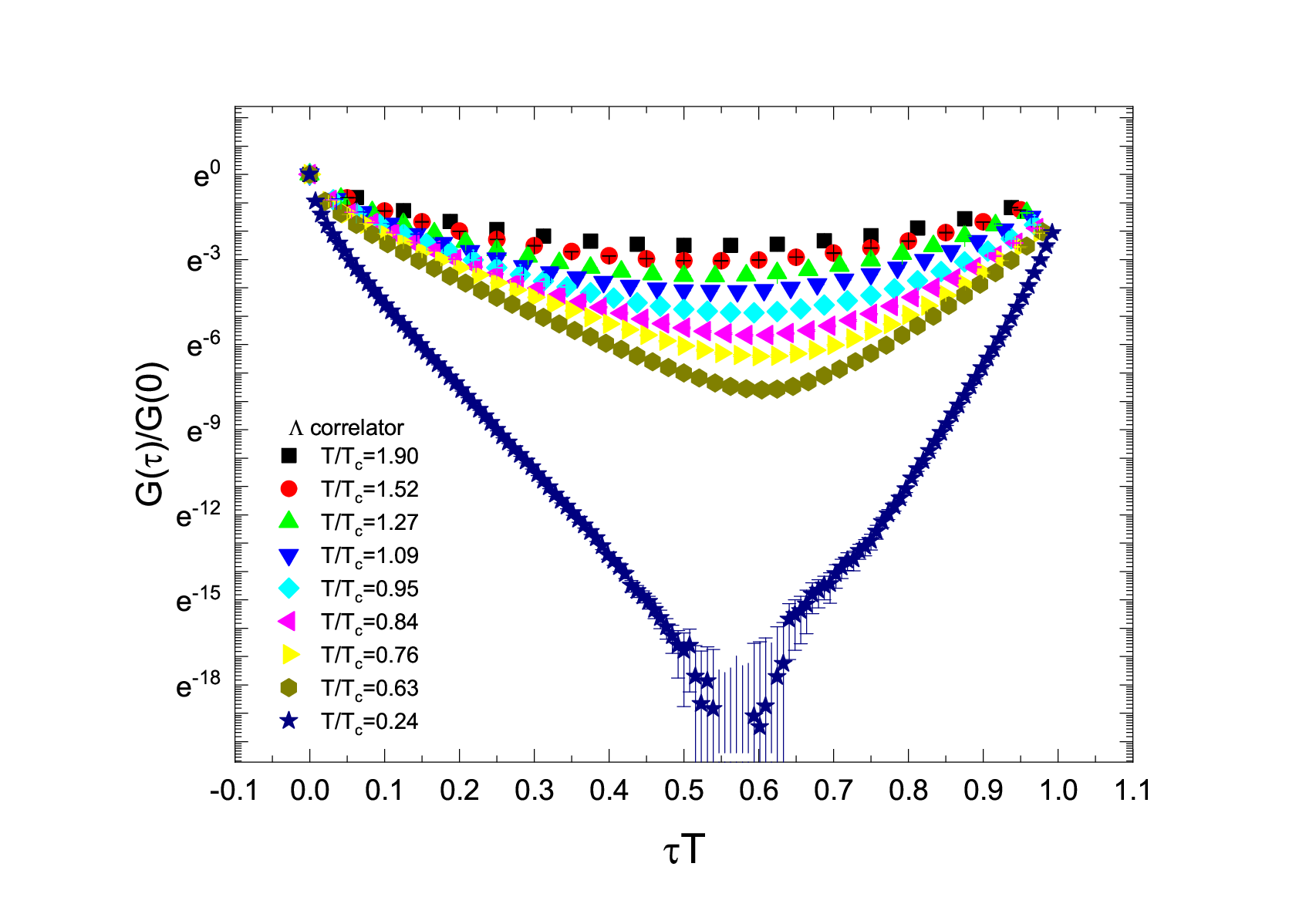}
		\caption{\label{fig1} Euclidean correlator $G(\tau)/G(0) $ of $\Lambda$ as a function of $\tau T$ at different temperatures. At the lowest temperature $T/T_c =0.24$, the correlators at some points are not displayed due to minus values.}
	\end{figure*}

	\begin{figure*}[t!]
		\includegraphics*[width=0.49\textwidth]{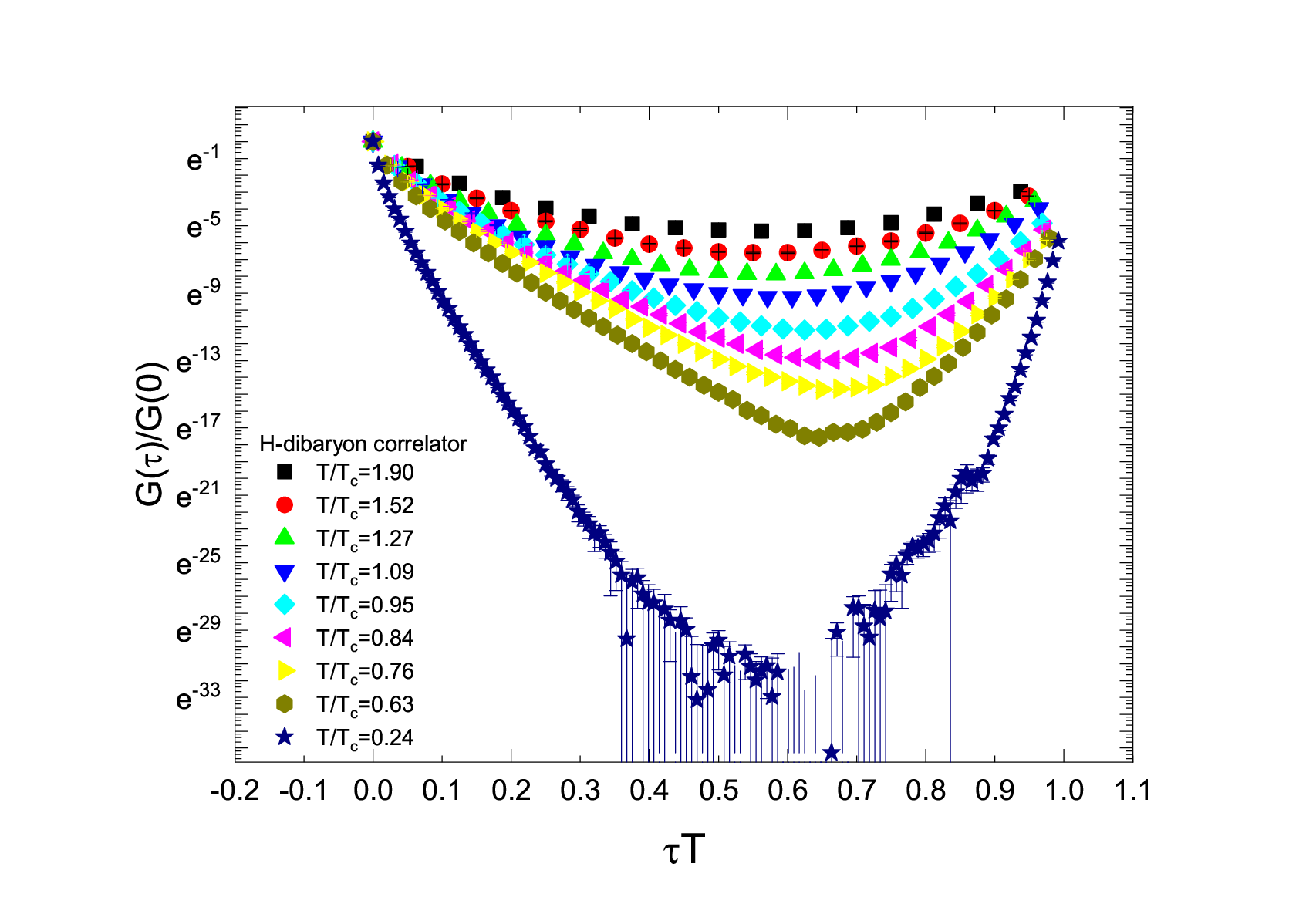}
		\caption{\label{fig2} Euclidean correlator $G(\tau)/G(0) $ of H-dibaryon as a function of $\tau T$ at different temperatures. At the lowest temperature $T/T_c =0.24$, the correlators at some points are not displayed due to  minus values.}
	\end{figure*}

\begin{figure}[t!]
		\centerline{\epsfxsize=3.0in\hspace*{0cm}\epsfbox{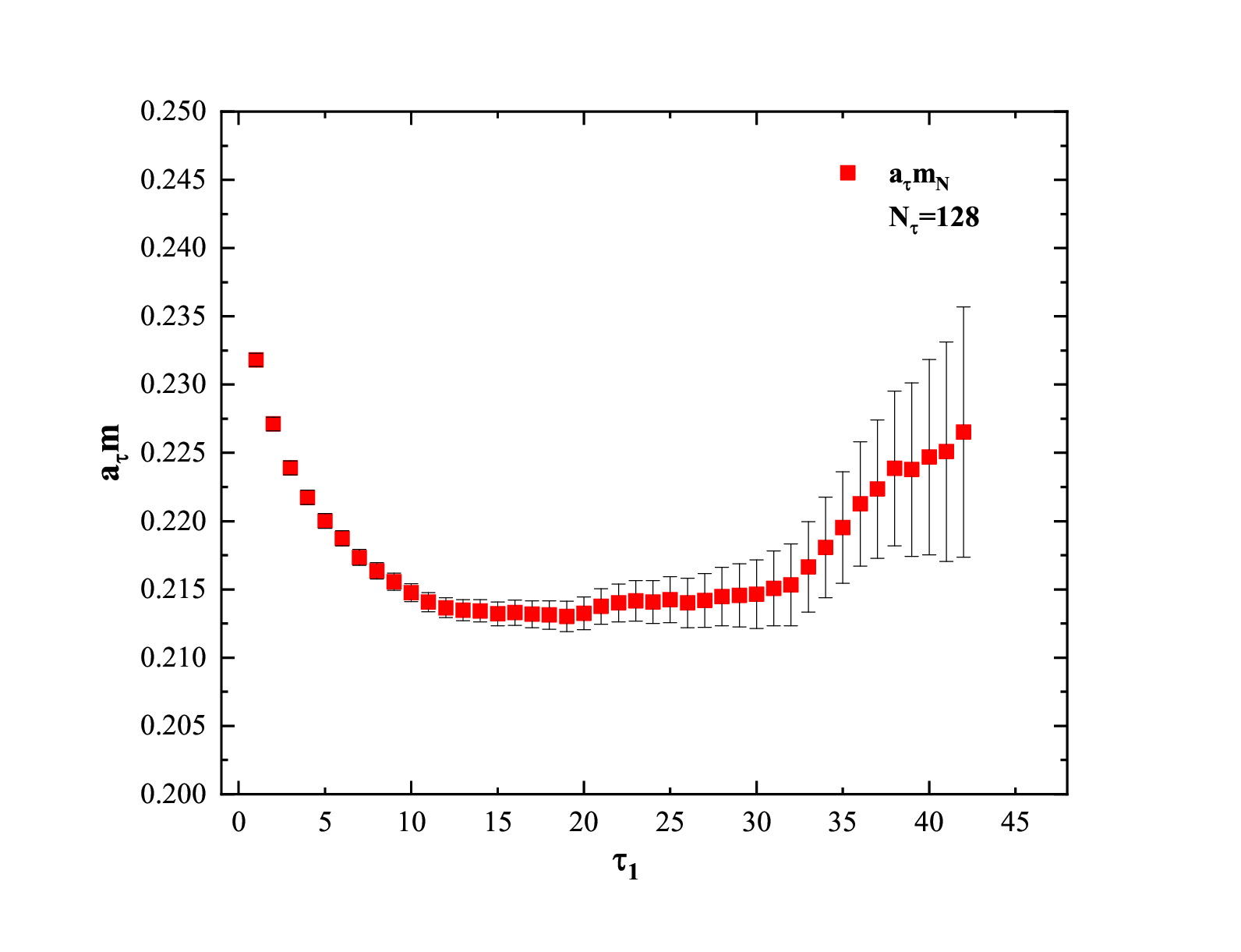}}%
		\centerline{\epsfxsize=3.0in\hspace*{0cm}\epsfbox{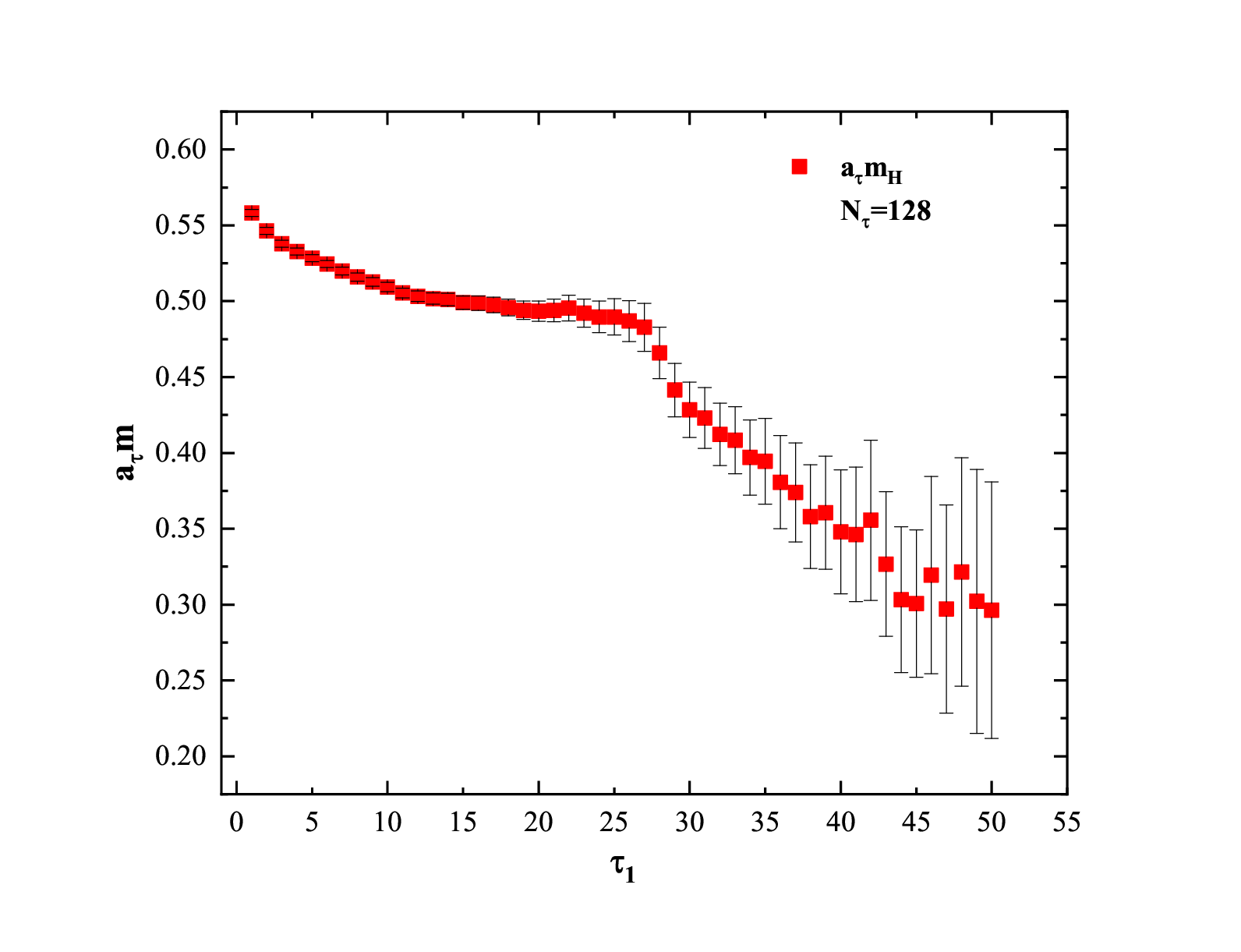}}
		\caption{\label{mass:suppressed} Mass values of nucleon and H-dibaryon obtained by fitting equation~(\ref{eq:Ansatz}) to correlators on $N_\tau=128 $ ensembles. The mass values are the fitting parameter $m_+$ in equation~\ref{eq:Ansatz} extracted by fitting procedure in different interval $[\tau_1, \tau_2]$.
Horizontal axis label $\tau_1$ represents different number of time slices suppressed which corresponds the lower bound of interval $[\tau_1, \tau_2]$. }
	\end{figure}

\begin{figure}[t!]
		\centerline{\epsfxsize=3.0in\hspace*{0cm}\epsfbox{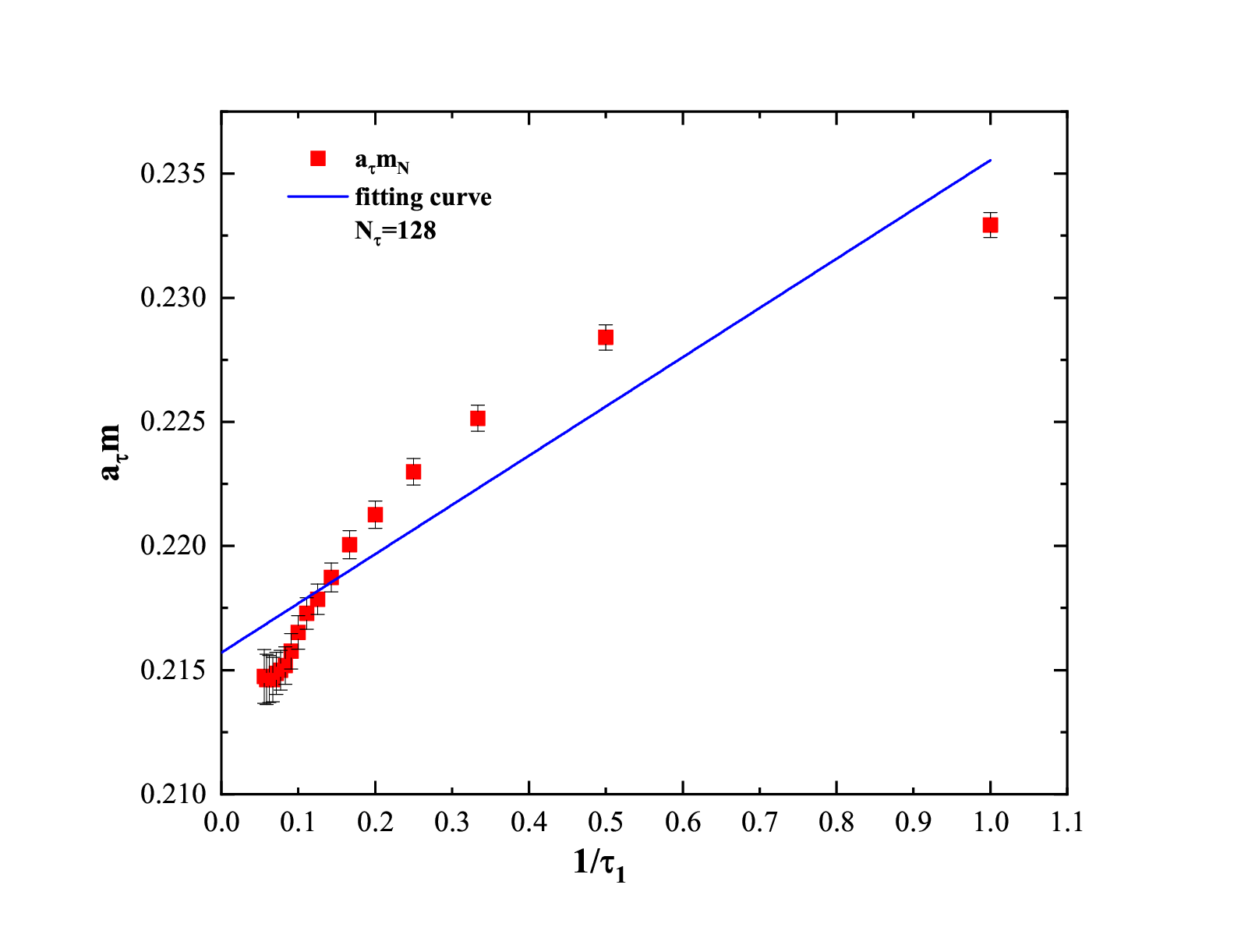}}%
		\centerline{\epsfxsize=3.0in\hspace*{0cm}\epsfbox{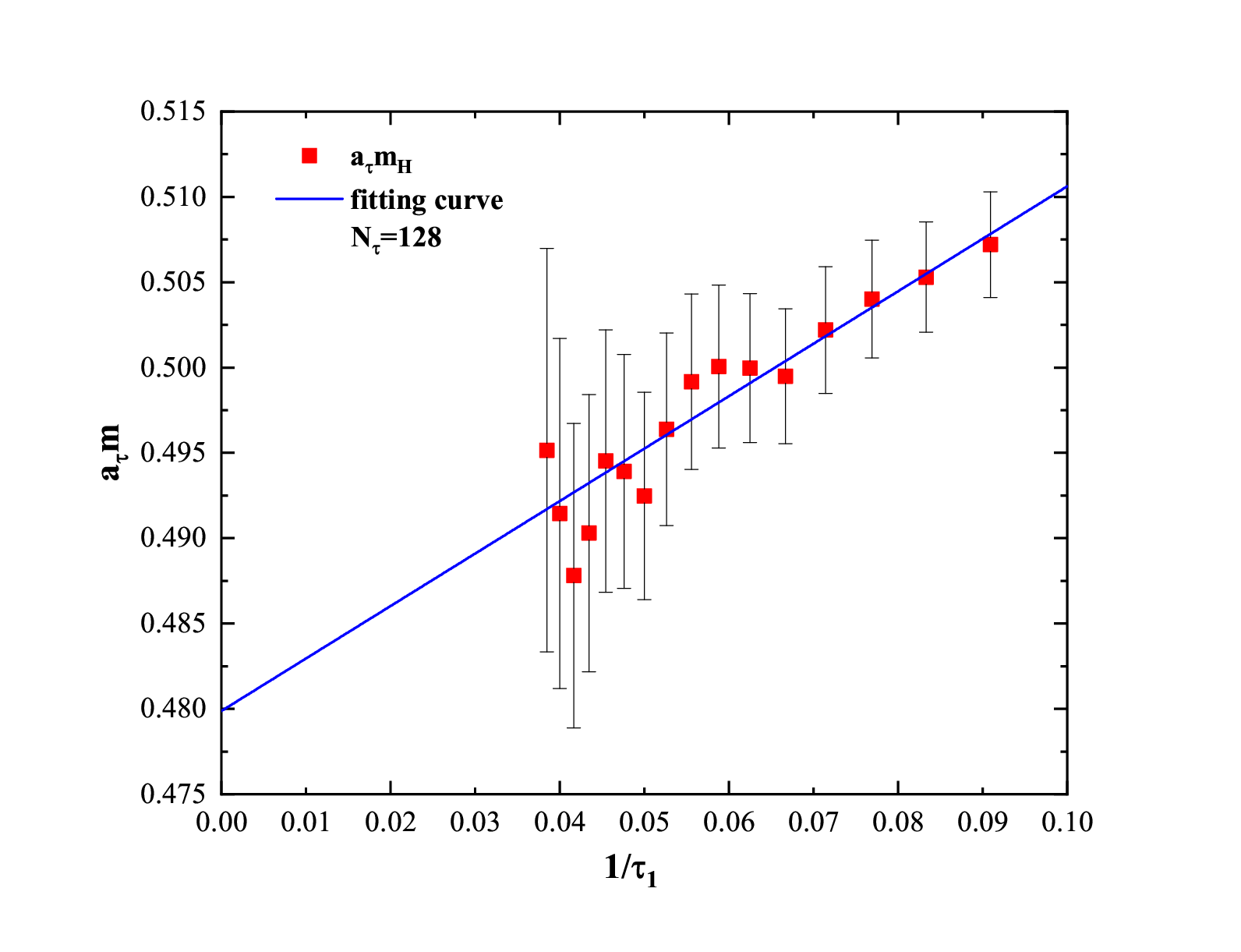}}
		\caption{\label{mass:extrapolation} Linear  extrapolation of mass values for nucleon and H-dibaryon on $N_\tau=128 $ ensembles. Horizontal axis represents inverse values of time slices suppressed. }
	\end{figure}

	\section{SPECTRAL FUNCTION}
	\label{SectionSpectral}
	Hadron properties are encoded in spectral functions which can provide us important information on hadrons. Two approaches and their variants are
  adopted to reconstruct spectral function. The first is the maximum entropy method and their variants~\cite{Asakawa:2000tr, Burnier:2013nla, Aarts:2007wj}.
 The second is the Backus-Gilbert
   method and their variants~\cite{Brandt:2015sxa,Brandt:2015aqk,Lupo:2023qna,Barone:2023iat}( reviews on the spectral function in lattice QCD can be found in
Ref.~\cite{Rothkopf:2022fyo,Meyer:2017ydp} and references therein ).  Recently,  based on the Backus-Gilbert
   method, a new method was presented in Ref.~\cite{Hansen:2019idp}. This method allows for choosing a smearing function at the beginning of the reconstruction procedure.
      To render this paper self-contained,   we  briefly present the method which was designed  in Ref.~\cite{Hansen:2019idp} in this section.
	In the following, the notations and symbols are almost the same as those used in Ref.~\cite{Hansen:2019idp}.

	The correlation function can be written as:
	\begin{align}
		\label{eq:spectral}
		G(\tau) = \int_0^\infty dE\rho_L(E)b(\tau,E),
	\end{align}
	with $\rho_L(E)$ being the spectral function.
	We choose the basis function as:
	\begin{align}
		b(\tau,E) = e^{-\tau} + e^{-(1/T-\tau)},
	\end{align}
	
	We can approximate $\rho_L(E_\star)$ by $\bar{\rho}_L(E_\star)$, where $\bar{\rho}_L(E_\star)$ can be evaluated by
	\begin{align}
		\bar{\rho}_L(E_\star) = \sum_{\tau=0}^{\tau_{m}} g_\tau(E_\star) G(\tau+1),
	\end{align}
	after those coefficients $g_\tau(E_\star)$ are determined.
	
	The coefficients $g_\tau(E_\star)$ are determined by minimizing the linear combination $W[\lambda,g]$ of  the deterministic functional $A[g] $ and error functional $B[g]$
	\begin{align}
		W[\lambda,g] =  (1-\lambda) A[g] + \lambda \frac{B[g]}{G(0)^2},
	\end{align}
	under the unit area constraint
	\begin{align}
		\int_0^\infty dE  \bar\Delta_\sigma (E,E_\star) =1,
	\end{align}
	where $A[g]$ is defined as:
	\begin{align}
		A[g] = \int_{E_0}^\infty dE | \bar\Delta_\sigma (E,E_\star) - \Delta_\sigma (E,E_\star)|^2,
	\end{align}
	with $\bar\Delta_\sigma (E,E_\star)$ and  $\Delta_\sigma (E,E_\star)$ being  the smearing function and target smearing function, respectively.
	These two functions are given by
	\begin{align}
		\bar\Delta_\sigma (E,E_\star)  = \sum_0^{\tau_m} g_\tau (\lambda, E_\star) b(\tau +1,E),
	\end{align}
	and
	\begin{align}
		\Delta_\sigma (E,E_\star)  = \frac{e^{-\frac{(E-E_\star)^2}{2\sigma^2}}} {\int_0^\infty dE e^{-\frac{(E-E_\star)^2}{2\sigma^2}}} ,
	\end{align}
	respectively.
	$B[g]$ is written as:
	\begin{align}
		B[g]= g^T {\rm Cov} g,
	\end{align}
	with ${ \rm Cov}$ is the covariance matrix of the correlation function $G(\tau)$. More details are given in Ref.~\cite{Hansen:2019idp}.

	\section{MC SIMULATION RESULTS}
	\label{SectionMC}
	
	Before presenting the simulation results, we describe the computation details.   The simulations are carried out on  $N_f=2+1$ Generation2 (Gen2)
	FASTSUM ensembles~\cite{Aarts:2020vyb} of which the ensembles at the lowest temperature are provided by the HadSpec
	collaboration~\cite{Edwards:2008ja,HadronSpectrum:2008xlg}, so the computation details are the same as those used in Ref.~\cite{Aarts:2020vyb}.
	We recompile the simulation details in the following three tables~\ref{tab:parameters},
	~\ref{tab:lattice_spacings}, and~\ref{tab:mass} from Ref.~\cite{Aarts:2020vyb}. 

The ensembles are generated with a Symanzik-improved gauge action and a
tadpole-improved clover fermion action, with stout-smeared links.
The details of the action are given in Ref.~\cite{Aarts:2020vyb}.  The parameters in the lattice action are recompiled in Table \ref{tab:parameters}.
The $N_f=2+1$ Gen2 ensembles correspond to a physical strange quark mass and a bare light quark mass of $a_\tau m_l=-0.0840$, yielding a pion mass
of $m_\pi=384(4)$ MeV (see Table \ref{tab:lattice_spacings}).

	\begin{figure}[h]
		\centerline{\epsfxsize=3.0in\hspace*{0cm}\epsfbox{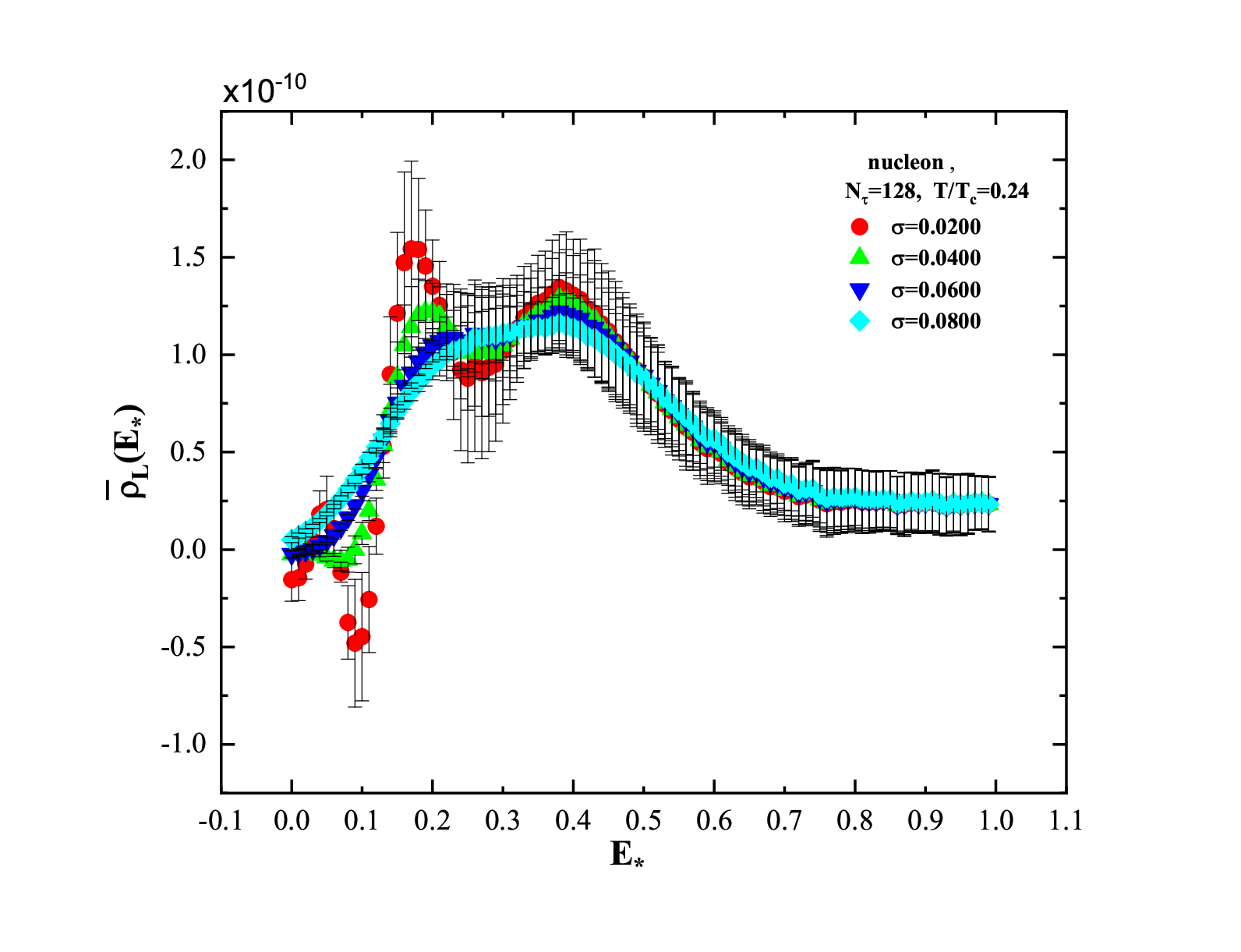}}%
		\centerline{\epsfxsize=3.0in\hspace*{0cm}\epsfbox{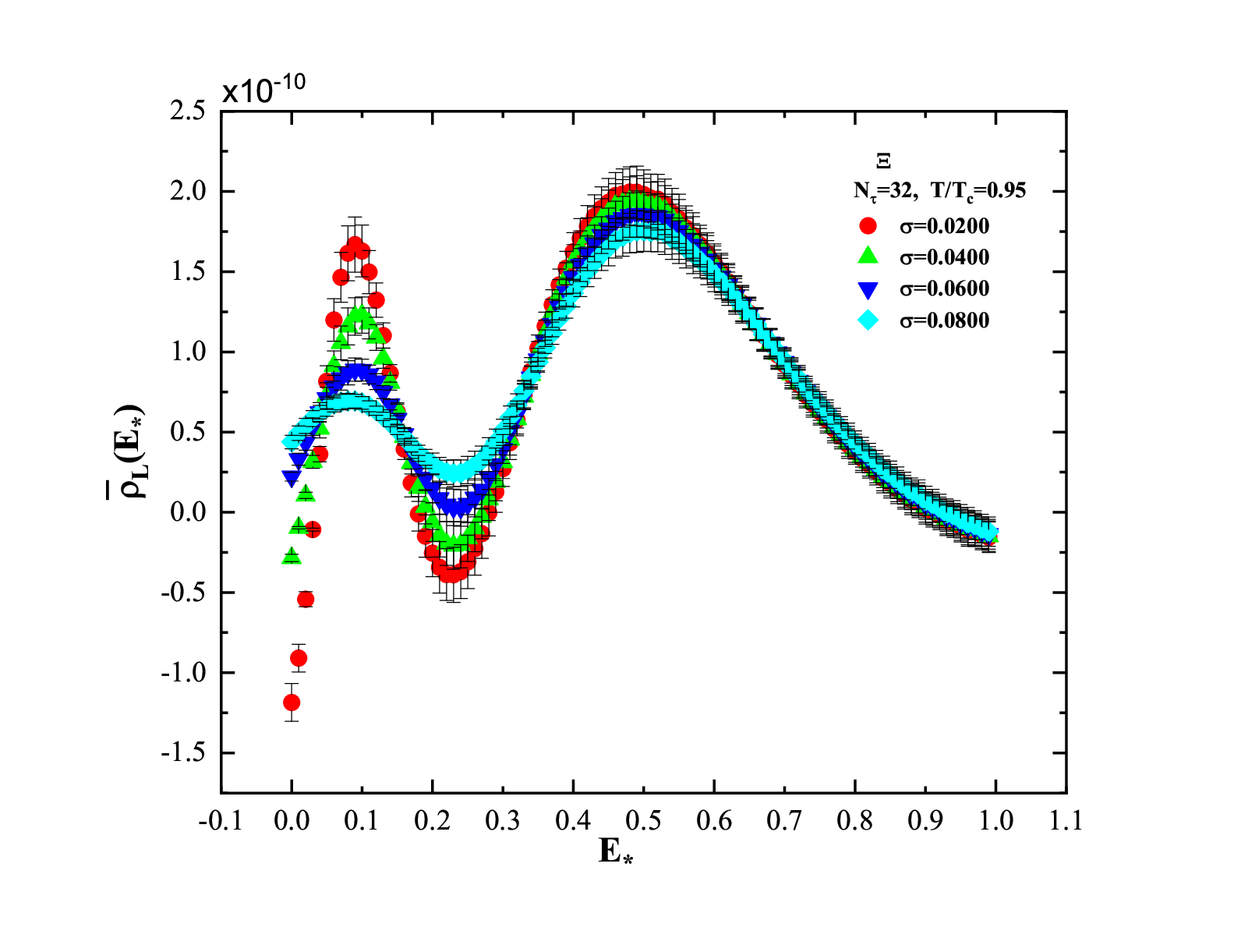}}
		\centerline{\epsfxsize=3.0in\hspace*{0cm}\epsfbox{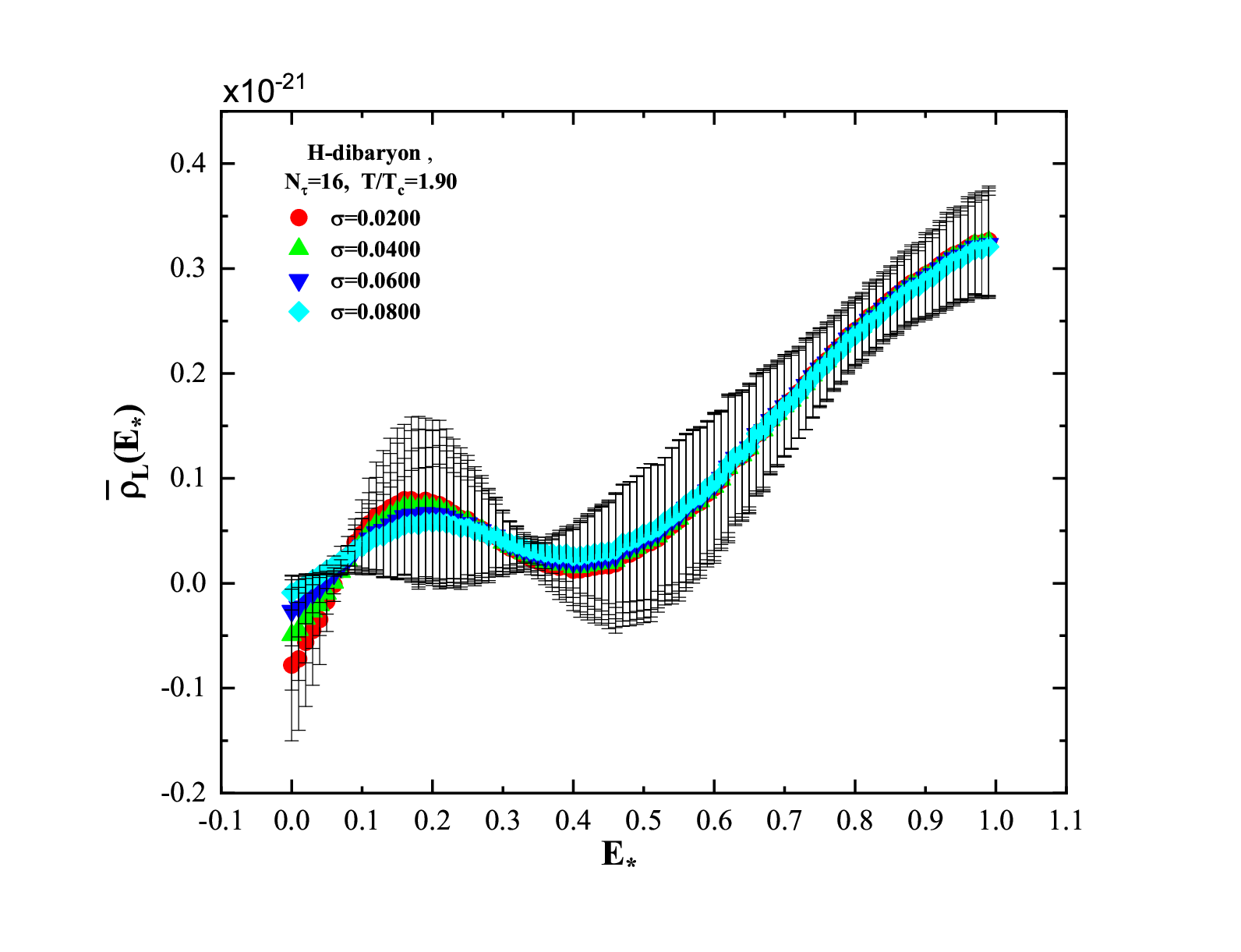}}
		\caption{\label{diff_s} Spectral density computed at different parameter $\sigma$ for the target smearing function $\Delta_\sigma (E,E_\star)$ for H-dibaryon, $\Xi$
			and $N$ at different temperatures.}
	\end{figure}

	\begin{figure}[h]
		\centerline{\epsfxsize=3.0in\hspace*{0cm}\epsfbox{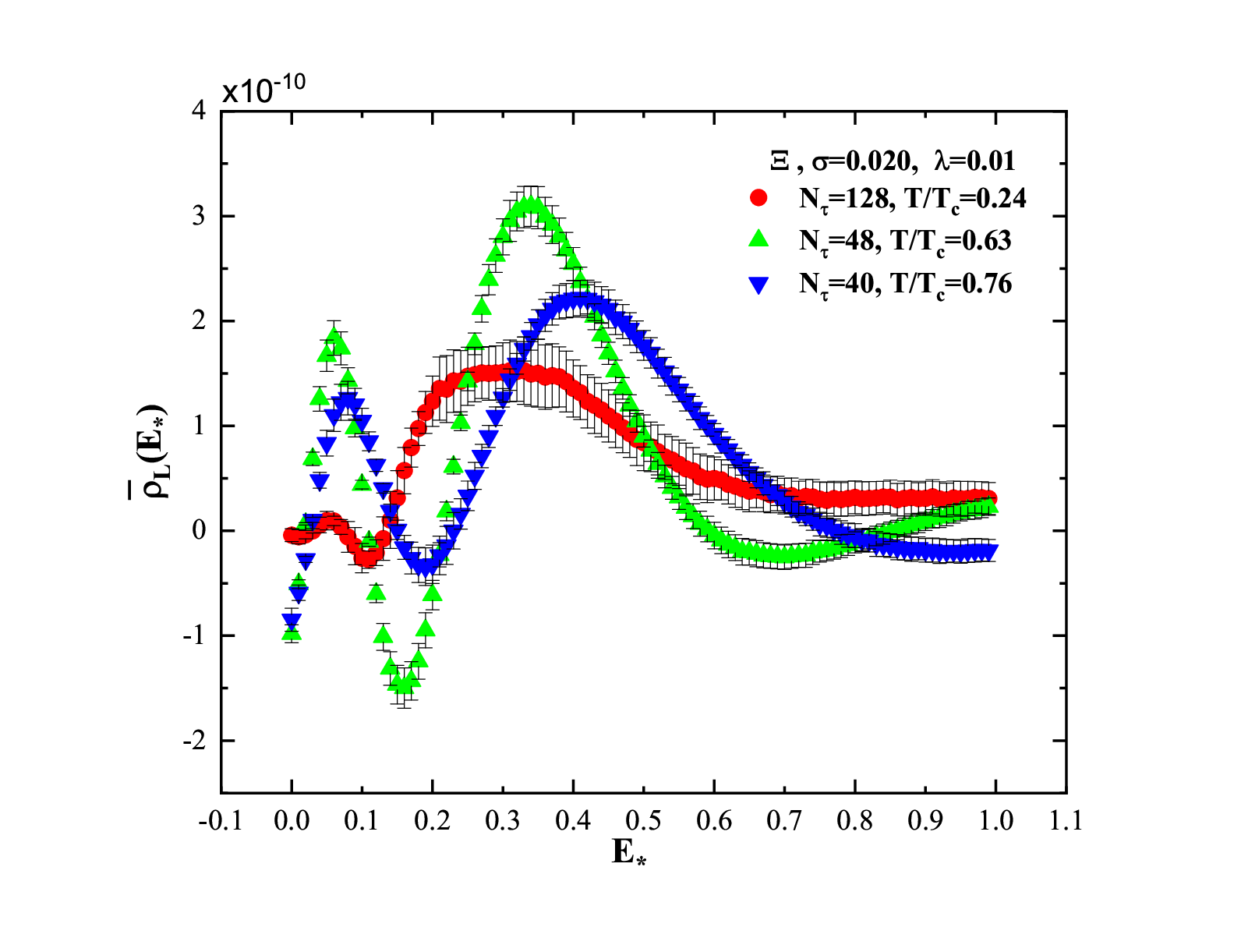}}%
		\centerline{\epsfxsize=3.0in\hspace*{0cm}\epsfbox{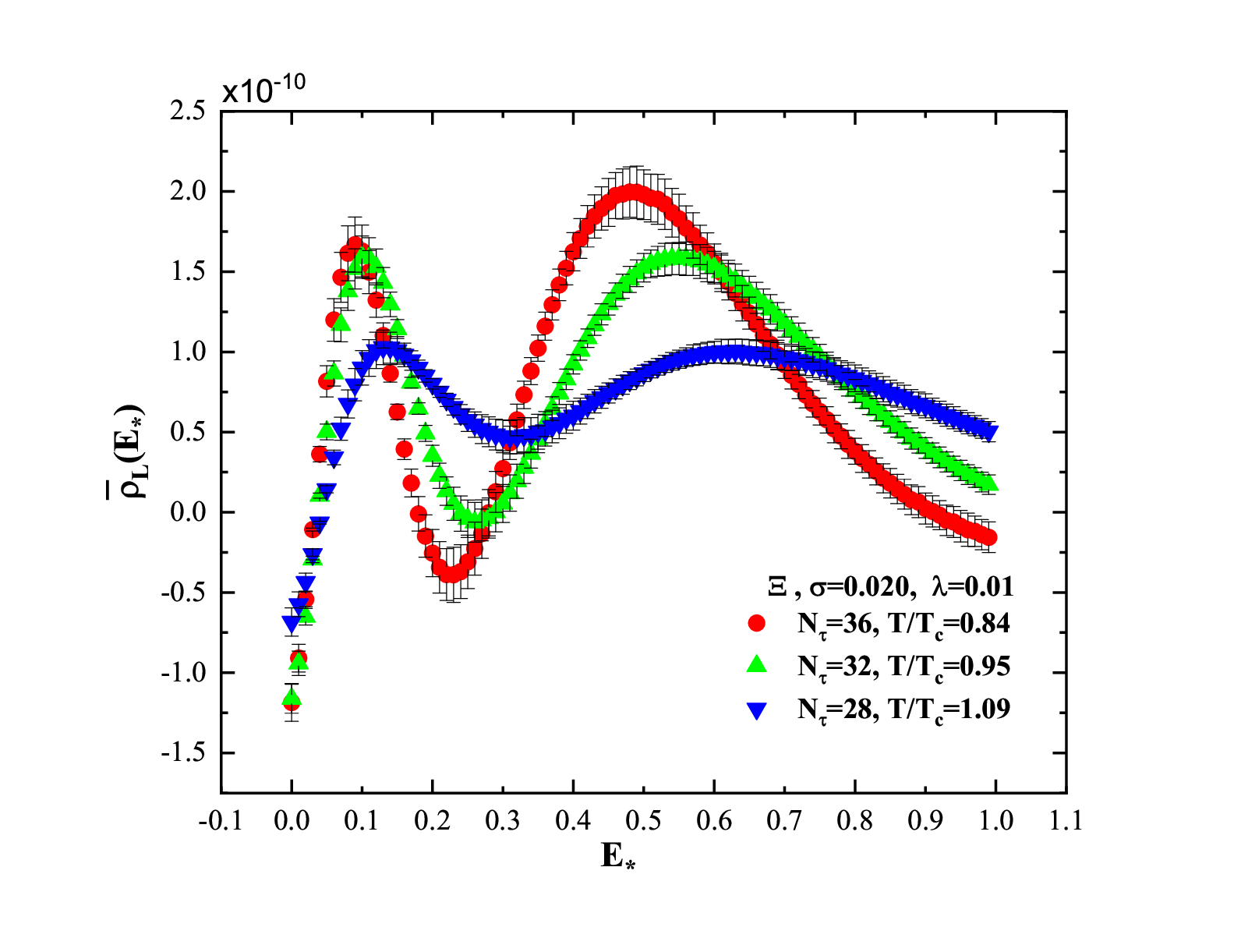}}
		\centerline{\epsfxsize=3.0in\hspace*{0cm}\epsfbox{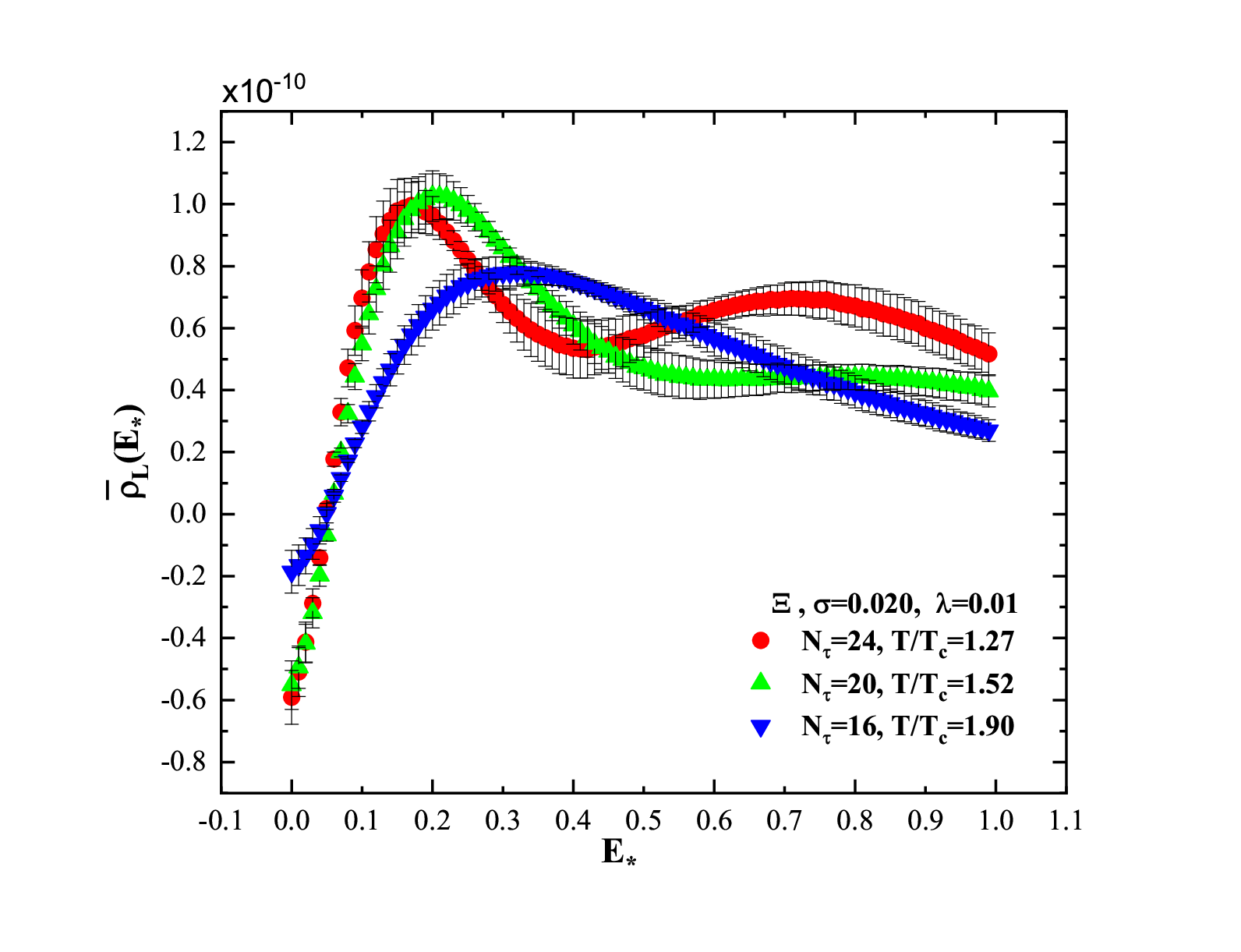}}
		\caption{\label{Xi_spectral} Spectral density of $\Xi$ at different temperatures.  }
	\end{figure}

	\begin{figure}[h]
		\centerline{\epsfxsize=3.0in\hspace*{0cm}\epsfbox{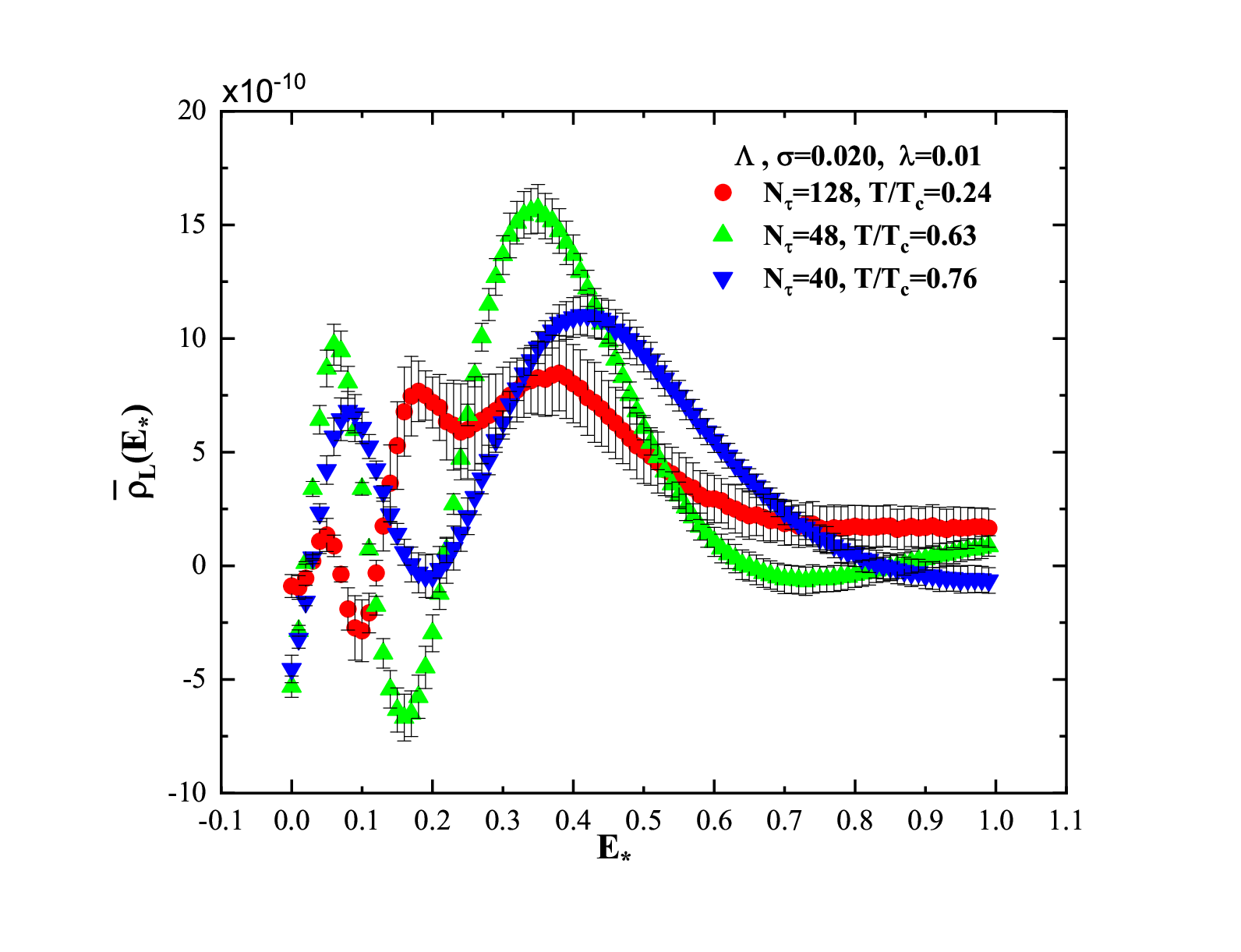}}%
		\centerline{\epsfxsize=3.0in\hspace*{0cm}\epsfbox{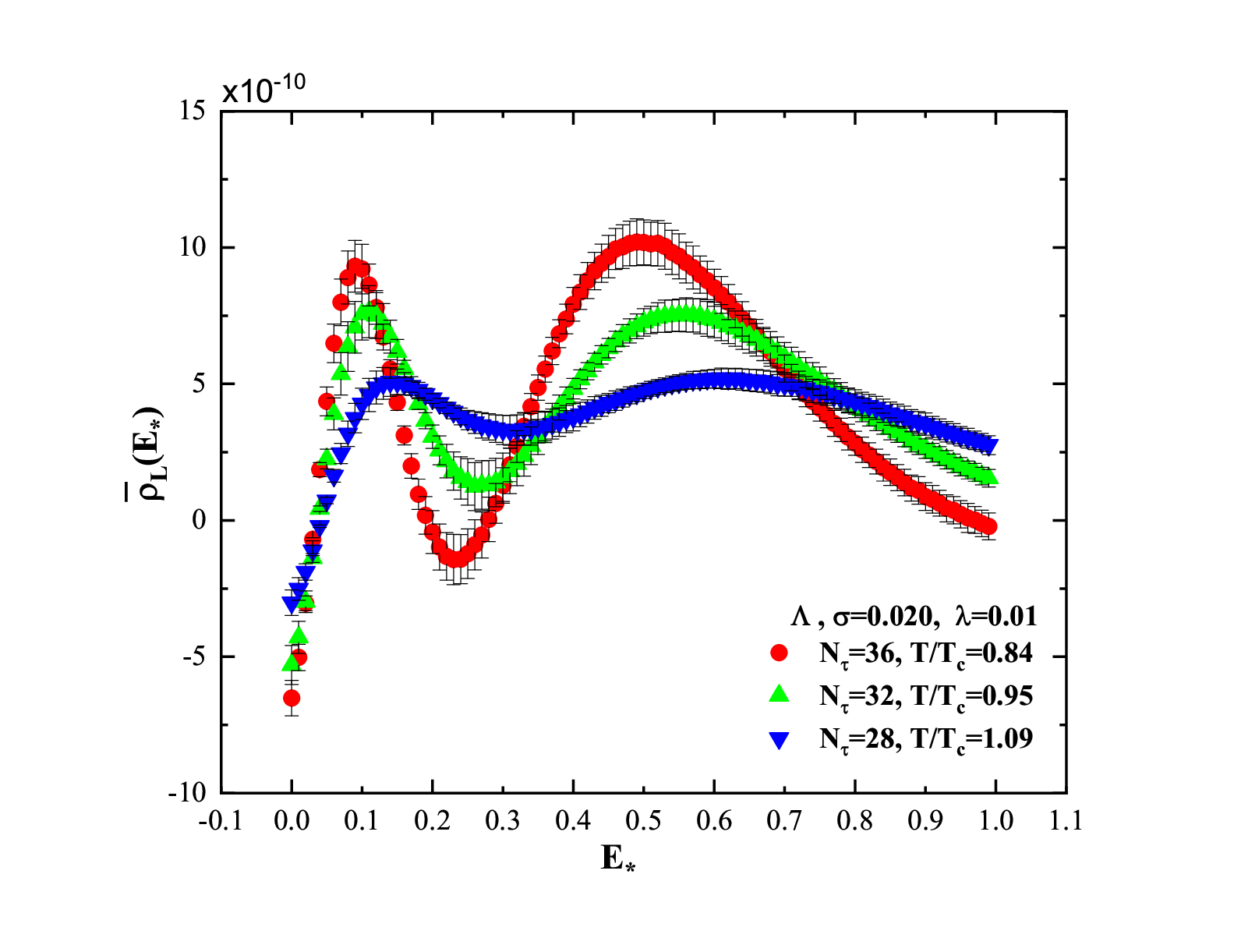}}
		\centerline{\epsfxsize=3.0in\hspace*{0cm}\epsfbox{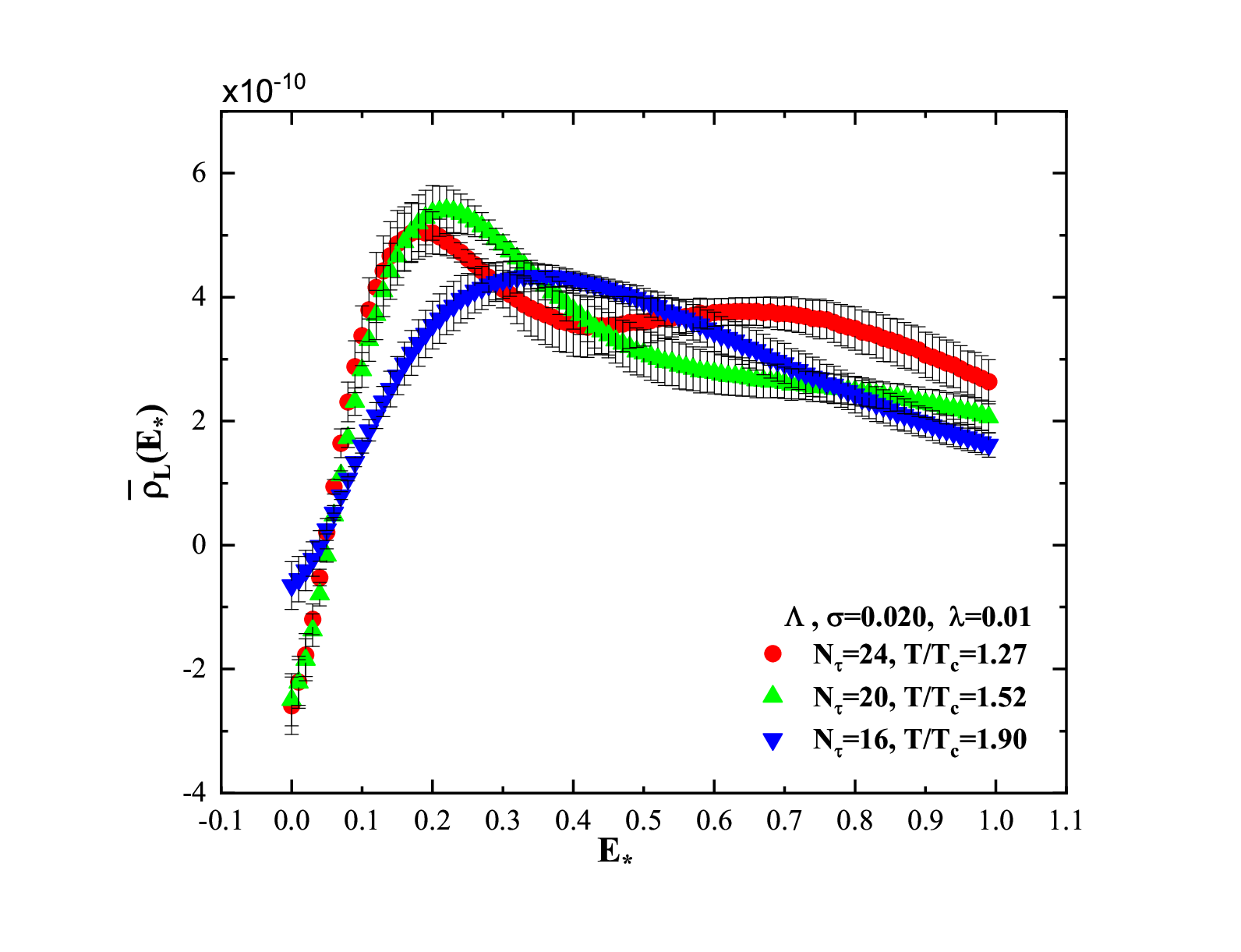}}
		\caption{\label{Lambda_spectral} Spectral density of $\Lambda$ at different temperatures. }
	\end{figure}

	\begin{figure}[h]
		\centerline{\epsfxsize=3.0in\hspace*{0cm}\epsfbox{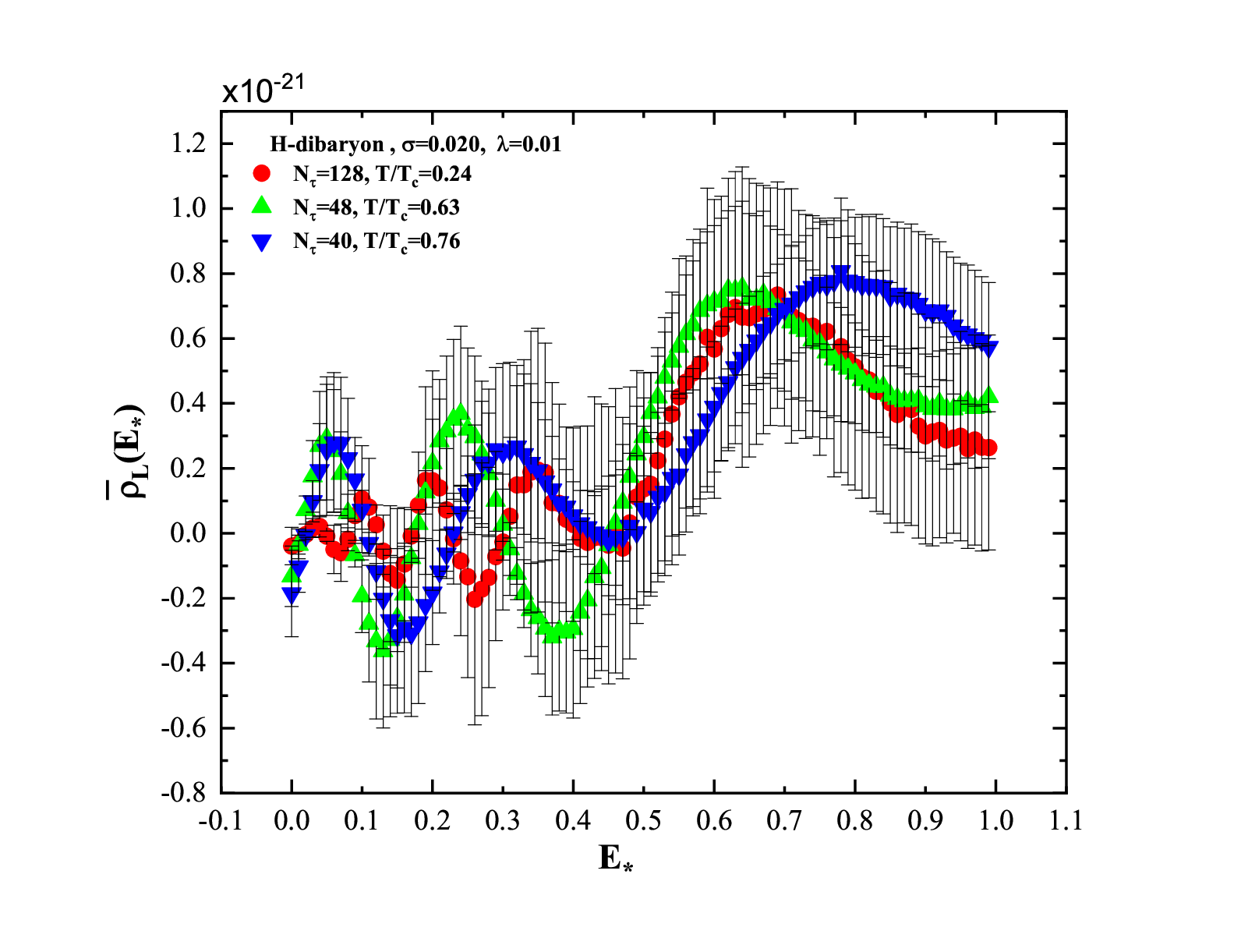}}%
		\centerline{\epsfxsize=3.0in\hspace*{0cm}\epsfbox{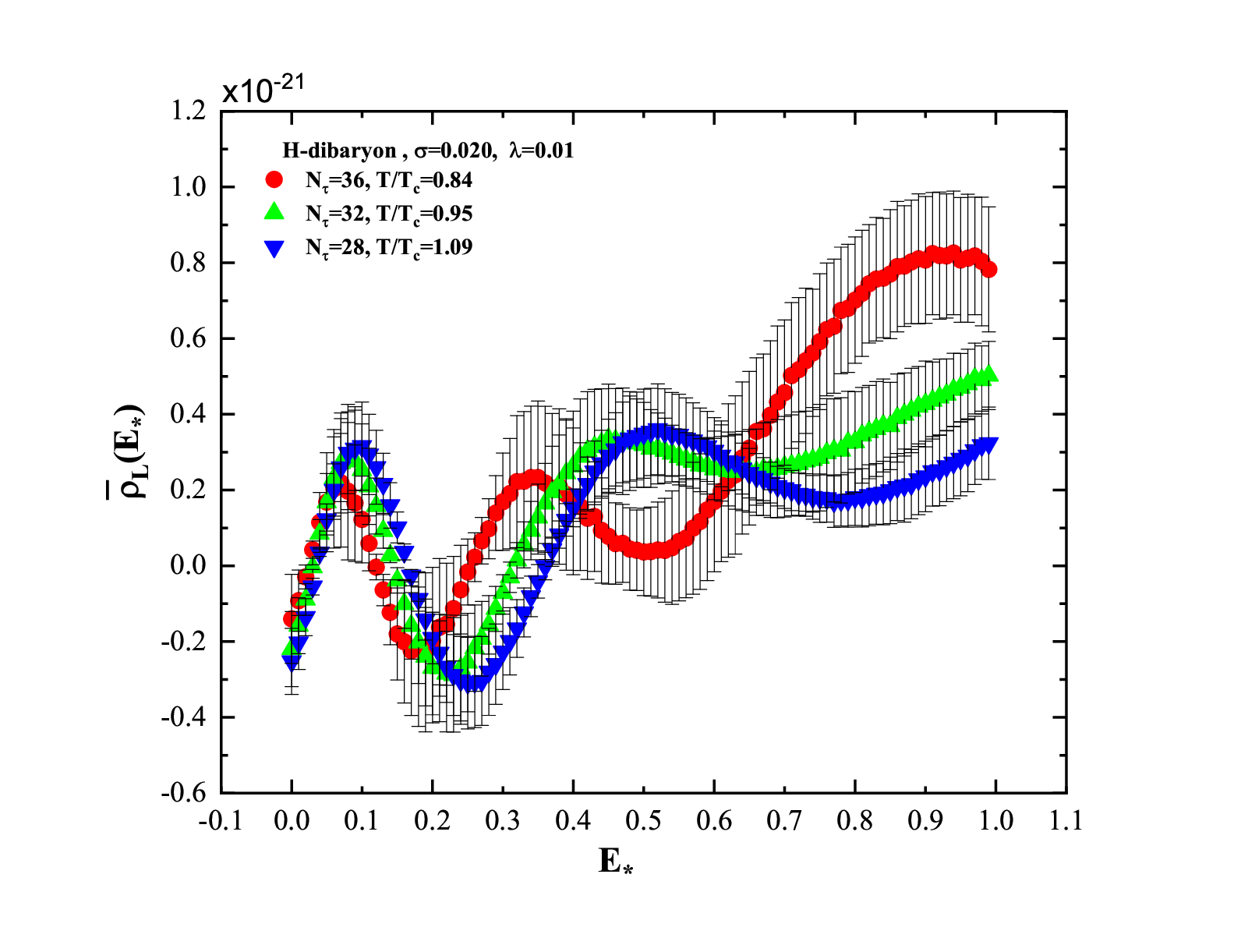}}
		\centerline{\epsfxsize=3.0in\hspace*{0cm}\epsfbox{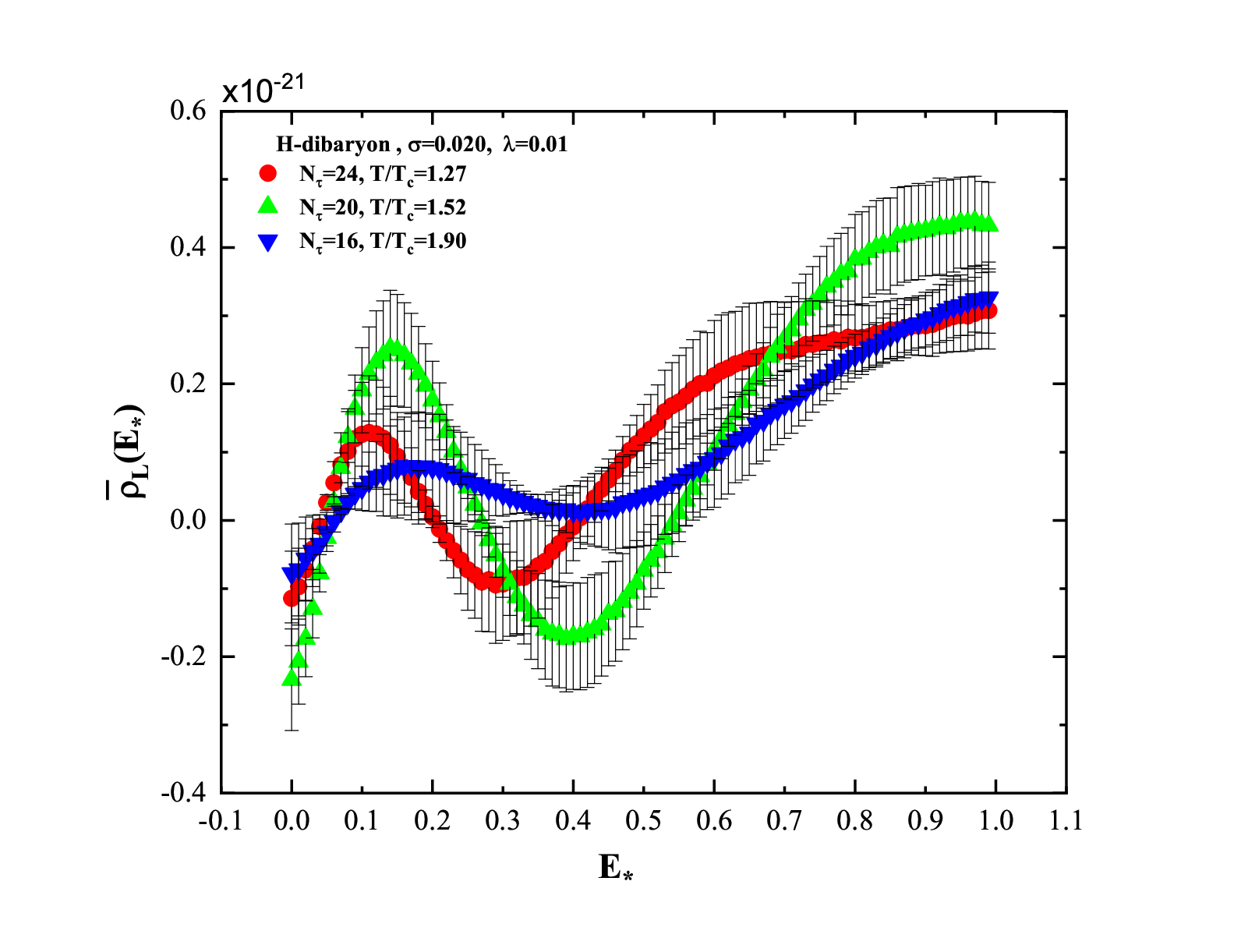}}
		\caption{\label{H_spectral} Spectral density of H-dibaryon at different temperatures. }
	\end{figure}
	
	\begin{figure*}[t!]
		\includegraphics*[width=0.49\textwidth]{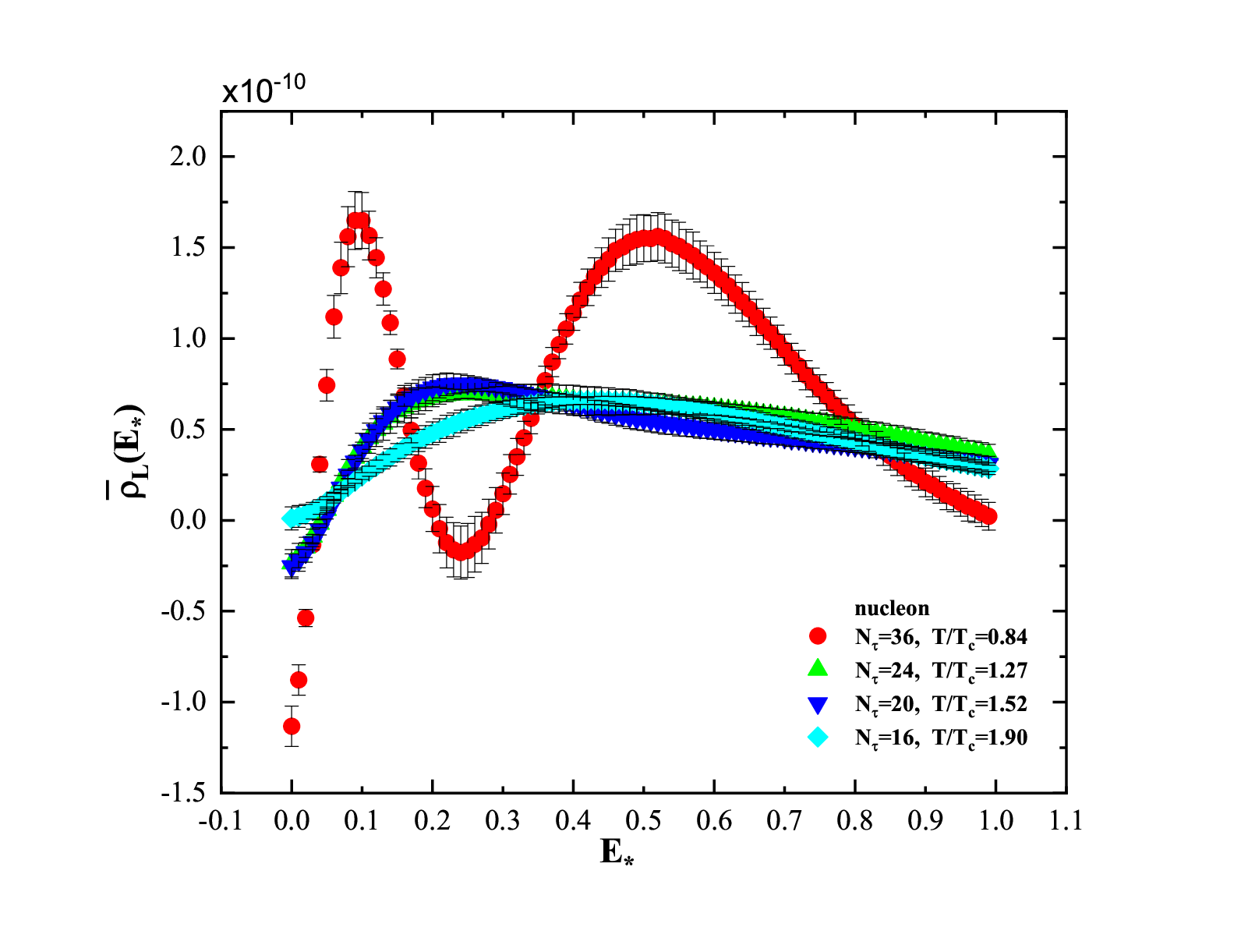}
		\caption{\label{fig21} Spectral density distribution of $N$ at different temperature $T/T_c = 0.84, 1.27, 1.52,1.90$.  At $T/T_c = 0.84, N_\tau = 36$,  the spectral density of $N$
			exhibits two peaks.  At $T/T_c = 1.27, 1.52,1.90$,  the spectral density distribution almost becomes smooth.}
	\end{figure*}
	
	\begin{figure*}[t!]
		\includegraphics*[width=0.49\textwidth]{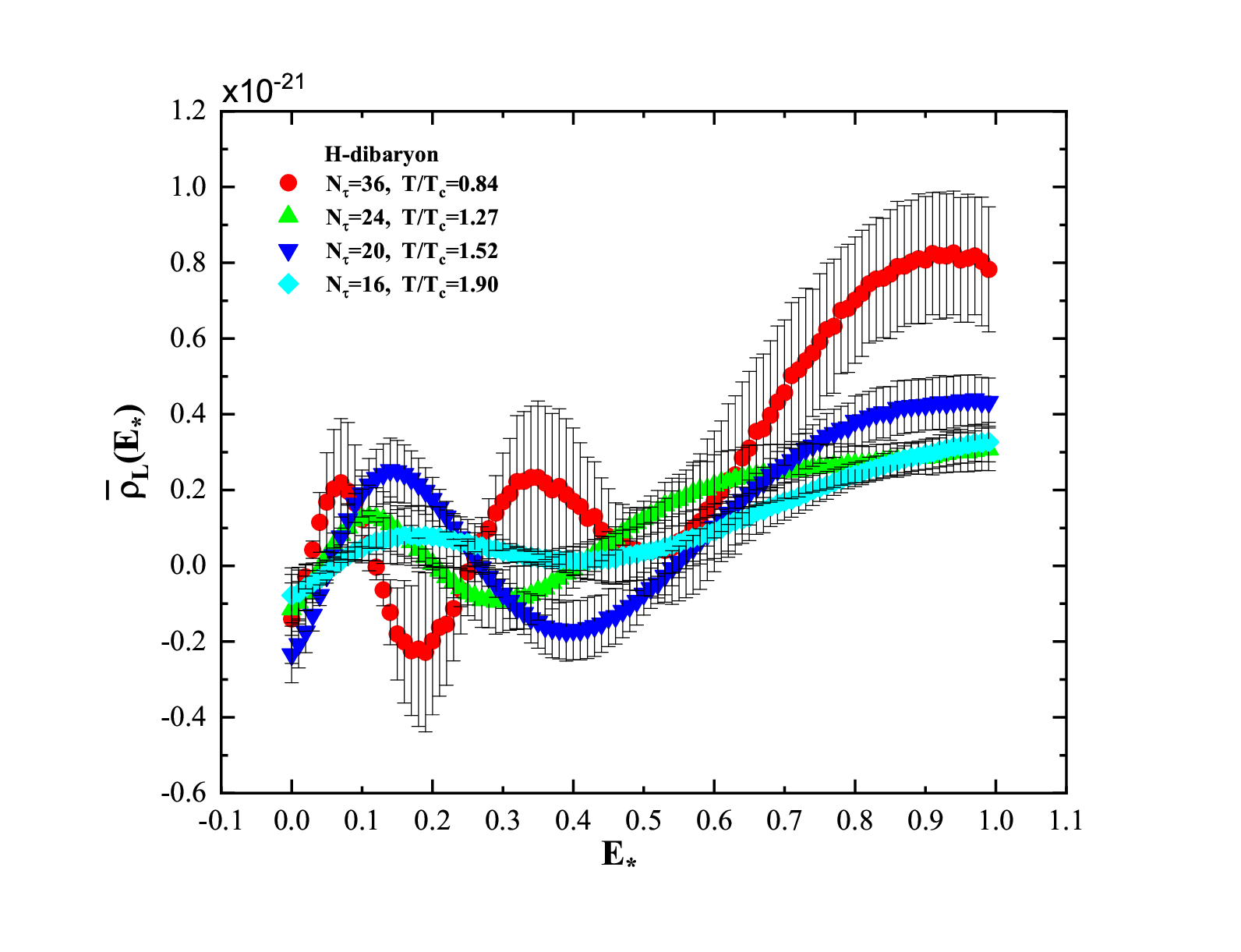}
		\caption{\label{fig22} Spectral density distribution of H-dibaryon at different temperature $T/T_c = 0.84, 1.27, 1.52,1.90$.  At $T/T_c = 0.84, N_\tau = 36$,  the spectral density of H-dibaryon
			exhibits two peaks.  At $T/T_c = 1.27, 1.52$,  the spectral density distribution has one peak. At $T/T_c = 1.90$,  the spectral density distribution almost becomes smooth. }
	\end{figure*}

\begin{table}[t]
	\caption{Parameters such as lattice spacing, pion mass etc collected from Ref.~\cite{Aarts:2020vyb} for Generation 2 ensemble.}
	\begin{ruledtabular}
		\begin{center}
			\begin{tabular}[t]{cc}
				$a_\tau$ [fm] & 0.0350(2)   \\
				\; $a_\tau^{-1}$ [GeV] \; & 5.63(4)  \\
				$\xi=a_s/a_\tau$ &  3.444(6)    \\
				$a_s$ [fm] & \; 0.1205(8) \; \;  \\
				$N_s$   &   24   \\
				$m_\pi$ [MeV] & 384(4)  \\
				$m_\pi L$ & 5.63  \\
			\end{tabular}
		\end{center}
		\label{tab:lattice_spacings}
	\end{ruledtabular}
\end{table}

The ensemble detail is listed in Table~\ref{tab:mass} which is  recompiled from Ref.~\cite{Aarts:2020vyb} with a slight difference on ensemble $N_s^3\times N_\tau = 32^3\times 48$.
The corresponding physical parameters such as the lattice spacing, and the pion mass etc are collected in Table ~\ref{tab:lattice_spacings}.
\begin{table*}[htp]
	\caption{Spatial and temporal extent, temperature in MeV, number of configurations,  mass of $N$, $\Sigma$, $\Xi$,  $\Lambda$ and H-dibaryon. Masses of baryons and H-dibaryon are obtained by
extrapolation method.
		Estimates of statistical and systematic errors are
		contained in the first and second brackets, respectively.  The errors of fitting parameters of linear extrapolation are taken to be the
systemtatic errors of the masses. The ensembles at the lowest temperatures were provided by HadSpec~\cite{Edwards:2008ja,HadronSpectrum:2008xlg} (Gen2).  }
	\begin{ruledtabular}
		\begin{center}
			\begin{tabular}[t]{ccccclllll}
				$N_s $ & \; $N_\tau$ & \; $T {\rm  [MeV]} $  & \;   $T/T_c$  & \;  $N_{\rm cfg}$  & \; $a_\tau m_N$ & \;  $a_\tau m_\Sigma$ & \;  $ a_\tau m_\Xi $ & \;  $a_\tau m_{\Lambda}$ & \; $a_\tau m_H $  \\
				\hline
			24 & 128 & 44 & 0.24 & 304  &  0.2133(24)(6)  &  0.2349(21)(5)  &  0.2459(18)(5)  &  0.2299(23)(6)  &  0.457(27)(5) \\
32 & 48 & 117 & 0.63 & 601  &  0.208(2)(3)  &  0.231(1)(2)  &  0.243(1)(2)  &  0.226(2)(3)  &  0.448(17)(4) \\
24 & 40 & 141 & 0.76 & 502  &  0.203(2)(5)  &  0.228(2)(4)  &  0.239(2)(4)  &  0.221(2)(4)  &  0.437(14)(9) \\
24 & 36 & 156 & 0.84 & 501  &  0.196(2)(6)  &  0.221(2)(5)  &  0.231(2)(5)  &  0.214(2)(5)  &  0.42(1)(2) \\
24 & 32 & 176 & 0.95 & 1000  &  0.181(2)(9)  &  0.204(2)(8)  &  0.215(2)(7)  &  0.199(2)(8)  &  0.393(7)(21) \\
24 & 28 & 201 & 1.09 & 1001  &  0.179(2)(12)  &  0.191(2)(12)  &  0.201(2)(11)  &  0.190(2)(11)  &  0.38(1)(2) \\
24 & 24 & 235 & 1.27 & 1001  &  0.172(3)(15)  &  0.179(3)(15)  &  0.191(3)(14)  &  0.182(3)(14)  &  0.36(1)(3) \\
24 & 20 & 281 & 1.52 & 1000  &  0.159(4)(18)  &  0.164(4)(18)  &  0.176(4)(17)  &  0.169(4)(17)  &  0.33(1)(4) \\
24 & 16 & 352 & 1.90 & 1000  &  0.154(6)(24)  &  0.158(6)(24)  &  0.171(6)(23)  &  0.164(6)(23)  &  0.31(2)(4) \\

			\end{tabular}
		\end{center}
		
		\label{tab:mass}
	\end{ruledtabular}
\end{table*}


\begin{table}[t]
	\caption{On different $N_\tau $ lattice, peak position $E_\star $  of spectral density for $N$.  }
	\begin{ruledtabular}
		\begin{center}
			\begin{tabular}[t]{c|cccc}
				
				& $N_\tau$  & $E_\star$ & $E_\star$ &  $E_\star$  \\
				\hline
				\multirow{9}{*}{$N$}    &  128  &   0.05   &   0.17   &   0.38    \\
				&  48  &   0.06   &   0.35   & -   \\
				&  40  &   0.09   &   0.42   & -   \\
				&  36  &   0.10   &   0.52   & -   \\
				&  32  &   0.11   &   0.56   & -   \\
				&  28  &   0.18   &   0.57   & -   \\
				&  24  &   0.26   & -  & -   \\
				&  20  &   0.24   & -  & -   \\
				&  16  &   0.43   & -  & -   \\						
			\end{tabular}
		\end{center}
		\label{tab:n}
	\end{ruledtabular}
\end{table}

\begin{table}[t]
	\caption{On different $N_\tau $ lattice, peak position $E_\star $  of spectral density for $\Sigma$. }
	\begin{ruledtabular}
		\begin{center}
			\begin{tabular}[t]{c|cccc}				
				& $N_\tau$  & $E_\star$ & $E_\star$ &  $E_\star$  \\
				\hline
				\multirow{9}{*}{$\Sigma$}    &  128  &   0.05   &   0.17   &   0.37  \\
				&  48  &   0.06   &   0.35   & - \\
				&  40  &   0.08   &   0.41   & - \\
				&  36  &   0.09   &   0.49   & - \\
				&  32  &   0.11   &   0.56   & - \\
				&  28  &   0.17   &   0.57   & - \\
				&  24  &   0.20   &   -   & - \\
				&  20  &   0.22   & -  & - \\
				&  16  &   0.39   & -  & - \\				
			\end{tabular}
		\end{center}
		\label{tab:s}
	\end{ruledtabular}
\end{table}

\begin{table}[t]
	\caption{On different $N_\tau $ lattice, peak position $E_\star $  of spectral density for $\Xi$. }
	\begin{ruledtabular}
		\begin{center}
			\begin{tabular}[t]{c|ccccc}				
				& $N_\tau$  & $E_\star$ & $E_\star$   \\
				\hline
				\multirow{9}{*}{$\Xi$}    &  128  &   0.05   &   0.33     \\
				&  48  &   0.06   &   0.34    \\
				&  40  &   0.08   &   0.41    \\
				&  36  &   0.09   &   0.48    \\
				&  32  &   0.10   &   0.55    \\
				&  28  &   0.13   &   0.63    \\
				&  24  &   0.17   &   0.71    \\
				&  20  &   0.20   &   -   \\
				&  16  &   0.32   & -   \\
		\end{tabular}
		\end{center}
		\label{tab:x}
	\end{ruledtabular}
\end{table}

\begin{table}[t]
	\caption{On different $N_\tau $ lattice, peak position $E_\star $  of spectral density for $\Lambda$. }
	\begin{ruledtabular}
		\begin{center}
			\begin{tabular}[t]{c|cccc}				
				& $N_\tau$  & $E_\star$ & $E_\star$ &  $E_\star$  \\
				\hline
				\multirow{9}{*}{$\Lambda$}    &  128  &   0.05   &   0.18   &   0.38  \\
				&  48  &   0.06   &   0.35   & - \\
				&  40  &   0.08   &   0.42   & - \\
				&  36  &   0.09   &   0.49   & - \\
				&  32  &   0.11   &   0.56   & - \\
				&  28  &   0.14   &   0.61   & - \\
				&  24  &   0.18   &   0.64   & - \\
				&  20  &   0.22   & -  & - \\
				&  16  &   0.35   & -  & - \\
			\end{tabular}
		\end{center}
		\label{tab:h}
	\end{ruledtabular}
\end{table}

\begin{table}[t]
	\caption{On different $N_\tau $ lattice, peak position $E_\star $  of spectral density for H-dibaryon. }
	\begin{ruledtabular}
		\begin{center}
			\begin{tabular}[t]{c|ccccc}				
				& $N_\tau$  & $E_\star$ & $E_\star$ &  $E_\star$ & $E_\star$  \\
				\hline
				\multirow{9}{*}{H-dibaryon}   &  128  &   0.10   &   0.19   &   0.35   &   0.69  \\
				&  48  &   0.05   &   0.24   &   0.64   & - \\
				&  40  &   0.06   &   0.32   &   0.78   & - \\
				&  36  &   0.07   &   0.35   & -  & - \\
				&  32  &   0.08   &   0.45   & -  & - \\
				&  28  &   0.10   &   0.52   & -  & - \\
				&  24  &   0.11   & -  & -  & - \\
				&  20  &   0.14   & -  & -  & - \\
				&  16  &   0.17   & -  & -  & - \\
			\end{tabular}
		\end{center}
		\label{tab:l}
	\end{ruledtabular}
\end{table}


The quark propagators are computed by using the deflation-accelerated algorithm~\cite{Luscher:2007es,Luscher:2007se}.  When computing the propagator,
The spatial links are stout smeared~\cite{Morningstar:2003gk} with two steps of smearing, using the weight $\rho = 0.14$. For the sources and sinks, we use the Gaussian smearing~\cite{Gusken:1989ad}
\begin{align}
	\eta' = C\left(1+\kappa H\right)^n\eta,
\end{align}
where $H$ is the spatial hopping part of the Dirac operator and $C$ an appropriate normalisation~\cite{Aarts:2015mma}.

The correlators of $\Lambda$ and H-dibaryon are presented in Fig.~\ref{fig1} and Fig.~\ref{fig2}, respectively.
For the correlators of $\Lambda$ and H-dibaryon, we find similar behavior which was displayed in Fig.~1 in Ref.~\cite{Aarts:2015mma} for   $N$.
For the correlator of $\Lambda$ on large $N_\tau$ and relatively small $N_s$ lattice, especially $24^3\times 128$ lattice,
some correlator data points are negative, and these points are not displayed on the plot, because the vertical axis is rescaled logarithmically.
At some points, the error bar looks strange, it is because at these points, the errors are the magnitude of the correlator value, and the
vertical axis is rescaled.  For the plot of H-dibaryon correlator, the same observation can be observed.

We use the extrapolation method to extract the ground state masses for  $N$, $\Xi$, $\Sigma$, $\Lambda$, and H-dibaryon. We first fit equation~(\ref{eq:Ansatz}) to correlator
by suppressing different early time slices  to get a series of mass values.  We present the results of nucleon and H-dibaryon on lattice $N_\tau=128$ in Fig.~\ref{mass:suppressed}.
After we get a series of mass values with different early time slices suppressed, we  extrapolate the mass values linearly with the scenario described in the last paragraph in Sec.~\ref{SectionLattice}. We present the results of linear extrapolation for nucleon and H-dibaryon on lattice $N_\tau=128$ in Fig.~\ref{mass:extrapolation}. In the extrapolation procedure,
we use one portion of the data presented in Fig.~\ref{mass:suppressed}.

The results are listed in table~\ref{tab:mass}.
We can find that the masses decrease when temperature increases.
We compare our results of
$N$ and $\Lambda $ below $T_c$ with those in Refs.~\cite{Aarts:2015mma,Aarts:2018glk}.  The results are consistent within errors.

We also calculate the spectral density $\bar\rho_L(E_\star)$  of the correlation function of $N$, $\Sigma$, $\Xi$, $\Lambda$ and H-dibaryon by using the public computer program~\cite{Hansen:code}.

We present the spectral density with different
$\sigma = 0.02, \ 0.04, \  0.06,\  0.08$ for $N$ , $\Xi$ and  H-dibaryon at three temperatures in Fig.~\ref{diff_s}.
The upper panel in Fig.~\ref{diff_s}
for $N$ at $T/T_c = 0.24 $ indicates that too large $\sigma$ value may skip peak structure of spectral density.
From the upper panel in Fig.~\ref{diff_s}, we can find that the spectral density distribution obtained by using $\sigma = 0.08$ has just one position where $\bar\rho_L(E_\star)$ takes locally maximum value.
The position is about at $E_\star =0.37$.  At $T/T_c =0.24$, the time extent $N_\tau=128$ is large enough to extract the ground state energy.

However, even if we do not suppress any early Euclidean time slices in the fitting procedure with equation~(\ref{eq:Ansatz}), we cannot get a mass value $a_\tau m_N$ which is larger than $0.30$.  The largest value of $a_\tau m_N $
we get by suppressing different number of early time slices is about $0.25$ which is smaller than 0.30. It can be seen clearly from Fig.~\ref{mass:suppressed}.
The mass value of 0.30 is somewhat an arbitrary value between the two peak positions of 0.17 and 0.38 for the spectral density from table~\ref{tab:n}.
So we think taking large $\sigma$ may lead to missing some peak structure. On the other hand, spectral density $\bar\rho_L(E_\star)$ obtained by using $\sigma = 0.02$  in the upper panel of Fig.~\ref{diff_s} has a peak position at
$E_\star \approx 0.05$  with small peak value. This peak structure may be due to the lattice artefact.

The middle panel in Fig.~\ref{diff_s} for $\Xi $  at $T/T_c = 0.95 $ shows that smaller $\sigma $ value can make  peak structure of
spectral density in small $E_\star $ region more pronounced.  The lower panel for H-dibaryon  at $T/T_c = 1.90 $ suggests  that different $\sigma $ value has little effect on the computation of
spectral density at high temperature.
So, we just present the spectral density results computed with $\sigma =0.020$ in the following.

The spectral density $\bar\rho_L(E_\star)$ of $\Xi$, $\Lambda$ and
H-dibaryon are given in Fig.~\ref{Xi_spectral}, ~\ref{Lambda_spectral} and ~\ref{H_spectral}, respectively.  The spectral density $\bar\rho_L(E_\star)$ of $N$ and $\Sigma$
has similar behaviour to that of $\Lambda$.
From Fig.~\ref{Xi_spectral} ,~\ref{Lambda_spectral} and ~\ref{H_spectral}, we can find that the spectral density
$\bar\rho_L(E_\star)$ of $\Xi$ and $\Lambda$ has similar behaviour,  while $\bar\rho_L(E_\star)$ of H-dibaryon is slightly different.
All the peak positions of  $\bar\rho_L(E_\star)$  are collected in
Table~\ref{tab:n},\ref{tab:s},\ref{tab:x},\ref{tab:l} and \ref{tab:h}.

From Fig.~\ref{Xi_spectral}, ~\ref{Lambda_spectral} and ~\ref{H_spectral},
we can find that at the lowest temperature $T/T_c = 0.24$, the spectral density for $\Xi$, $\Lambda$ and H-dibaryon has rich peak structure. Despite there are two peaks approximately
between $E_\star=0.20$ and
$E_\star=0.40$,   the spectral density $\bar\rho_L(E_\star)$ of $\Xi$ and $\Lambda$ in the range of $E_\star$ from  $0.20$ to
$0.40$ are almost the same.  The mass values  $a_\tau m_\Xi = 0.2459 $,
$a_\tau m_\Lambda = 0.2299 $ obtained by extrapolation method are in that range of  $E_\star $.

However,  $a_\tau m_H = 0.44 $ at $T/T_c = 0.24$ for H-dibaryon is in the neighbour of peak position $E_\star = 0.35$ where the $\bar\rho_L(E_\star)$  value is not very large.
Obtaining $a_\tau m_H = 0.44 $ at $T/T_c = 0.24$ is just by suppressing more early Euclidean time slices.
From the
upper panel of Fig.~\ref{H_spectral}, more high frequency components of the spectral density should be suppressed in the extrapolation procedure.

When temperature increases, the multi-peak structure of spectral density distribution
turns into two-peak structure for $\Xi$ and $\Lambda$ until at high temperature $T/T_c = 1.27, 1.52, 1.90$, the  spectral density distribution has one peak.

At the intermediate temperatures, the spectral density $\bar\rho_L(E_\star)$ has a two-peak structure. If we take the smaller values of $E_\star$ at peak positions as the ground
state energies of corresponding particle, then these mass values obtained by  the peak position of $\bar\rho_L(E_\star)$  are smaller than those mass values obtained in Ref.~\cite{Aarts:2015mma} and Ref.~\cite{Aarts:2018glk}.
Mass values of $a_\tau m_\Xi  $, $a_\tau m_\Lambda  $
and $a_\tau m_H $ presented in Table.~\ref{tab:mass}  are not consistent with the peak positions of corresponding spectral density.
This observation shows that the mass values obtained
by extrapolation method are affected by the two-peak structure of spectral density.

At high temperature $T/T_c = 1.27, 1.52, 1.90$, for the spectral density distribution for $N$,   the  spectral density exhibits one peak structure, and
the peak position shifts towards large value with increasing temperature.  In the meantime,
we can find the peak broadens and becomes smooth. It means that in the mass spectrum structure of nucleon, there is no $\delta$ function structure
contributing to the correlation function.
  We can find this observation from Fig.~\ref{fig21}
for $N$.
Similar behaviour can be found for $\Sigma$, $\Xi$ and $\Lambda$.
We think the smooth distribution of spectral density implies that there does not exist one-particle state at high temperature.

This is not the case for H-dibaryon. The spectral density distribution for H-dibaryon at $T/T_c = 1.27, 1.52, 1.90$ is presented in Fig.~\ref{fig22} from which we can find that at
$T/T_c = 1.27, 1.52$, the spectral density distribution still exhibits one peak structure until at $T/T_c = 1.90$, the spectral density distribution broadens and  becomes smooth. This observation may imply
that at temperature $T/T_c = 1.27, 1.52$,  H-dibaryon still remains as a one-particle state.

\section{DISCUSSIONS}\label{SectionDiscussion}

We have made a simulation in an attempt to determine the masses of the conjectured  H-dibaryon with $2+1$ flavor QCD with clover fermion
at nine different temperatures.
In the meantime, we also calculate the masses of $N$, $\Sigma$, $\Xi$ and $\Lambda$. The results  are collected in table~\ref{tab:mass}.  The
spectral density distribution of those particle's correlation function are computed to understand the mass spectrum obtained by extrapolation method.

In our simulation, the change of temperature is represented by the change of $T/T_c$. $T_c$ is pseudocritical temperature determined via renormalized
Polyakov loop and estimated to be $T_c = 185(4)\, {\rm MeV}$~\cite{Aarts:2014nba,Aarts:2020vyb}.

We have compared two scenarios to obtain the ground state mass as  possible as we can. One is to suppress more early Euclidean time slices in the fitting procedure
with equation~(\ref{eq:Ansatz}).
The second method is the extrapolation method. We  extrapolate some fitting results which are obtained with different Euclidean time slices suppression to time approaching infinity.
The results by the two methods are consistent with each other within errors. We just present the results by the extrapolation method in table~\ref{tab:mass}. In fact, among the series
of mass values obtained by different early time suppression, it is difficult to choose which mass value is the proper one. However, the extrapolation method can alleviate this difficulty
to some extent.

The analysis of spectral density can provide insights on mass spectrum. However, in our simulation, we can find that the mass values obtained by
extrapolation method are not consistent with the peak position of spectral density in some situations, especially at high temperature.  Under such situations, we think the results of extrapolation
method are more reliable. We can take the peak positions for nucleon in Table~\ref{tab:n} as an example to give an explanation.
The smaller values of peak position are increasing with temperature. However, with increasing temperature, the mass values of particle concerned are supposed to decrease.

At the lowest temperature $T/T_c = 0.24$, the spectral density distribution has rich peak structure.
The mass spectrum of particles  approximately reflects the peak position of spectral density distribution.

At the intermediate temperatures, the spectral density distribution exhibits a two-peak structure. The peak structure at the
smaller $E_\star$ becomes smooth gradually when $\sigma$ increases. Considering the quark mass which corresponds to $m_\pi=384(4){\rm MeV}$,  if we take
the smaller $E_\star$ at the peak position as the ground state energy, the mass value is too small.
So we think the mass values obtained by using extrapolation method
are affected by the two states.

At high temperature, despite we get the mass values for $N$, $\Sigma$, $\Xi$ and $\Lambda$ which
are presented in table~\ref{tab:mass} by extrapolation method,
the spectral density distribution appears to become smooth which implies
there does not exist one state.

H-dibaryon is a multi-baryon state. From the spectral density distribution, it is found that at the lowest temperature $T/T_c = 0.24$, the  multi-state structure manifests.
When temperature increases, the number of peaks decreases until at $T/T_c = 1.90 $, the spectral density distribution becomes almost smooth. It means it is likely that H-dibaryon survives beyond $T_c$ until it melts down at $T/T_c = 1.90 $. Considering that H-dibaryon is a multi-baryon state, this conclusion awaits further investigation.

It is appropriate to consider the lowest temperature ensembles $N_\tau=128$ to be the zero temperature ones since $N_\tau>\xi N_s$~\cite{Aarts:2020vyb}.
Using the mass values of H-dibaryon and $\Lambda$ in Table~\ref{tab:mass} at $N_\tau=128, T/T_c = 0.24$, an estimation of $\Delta m = m_H - 2m_\Lambda  $ can be made.
$\Delta m = m_H - 2m_\Lambda = -0.0026(11)  $ which is converted into physical unit to be  $\Delta m = m_H - 2m_\Lambda = -14.6(6.2) {\rm MeV} $.
Ref.~\cite{Beane:2006mx,NPLQCD:2012mex,Inoue:2010es,
	Inoue:2011ai,Francis:2018qch,Green:2021qol} showed the presence of a binding state of H-dibaryon, despite the disagreement on the binding energy values.
Ref.~\cite{Beane:2006mx,NPLQCD:2012mex} reported a binding energy of $16.6 \pm 2.1 \ {\rm MeV}$ at $m_\pi= 389\ {\rm MeV}$, and $74.6 \pm 4.7 \ {\rm MeV}$
at $m_\pi= 800\ {\rm MeV}$.
Ref.~\cite{Inoue:2010es} obtained a  binding energy of H-dibaryon $30-40\ {\rm MeV}$ for pion mass $673-1015\ {\rm MeV}$. Ref.~\cite{
	Inoue:2011ai} had the similar results.
Ref.~\cite{Francis:2018qch} published the binding energy $19 \pm 10 \ {\rm MeV} $ of H-dibaryon for $m_\pi = 960\  {\rm MeV}$. Ref.~\cite{Green:2021qol} presented an
estimation of the binding energy $4.56 \pm 1.13 \ {\rm MeV}$ in the continuum limit at the SU(3)-symmetric point
with $m_\pi = m_K \approx 420 \ {\rm MeV}$ ( for the binding energy versus pion mass, see also, Fig. 5 in Ref.~\cite{Green:2021qol} ).

In our simulation, the correlators of proton, $\Lambda$ and H-dibaryon at $N_\tau = 128$ have negative values, and in the fitting process for H-dibaryon,
we drop the negative values. We guess the emergence of negative values of the correlators is due to deterioration of the signal-to-noise ratio.

Our simualtions are at $m_\pi = 384(4) \ {\rm MeV}$ which is far from physical pion mass, so simulations with lower pion mass are expected to
give us more information about the properties of H-dibaryon.

\begin{acknowledgments}
	We thank Gert Aarts, Simon Hands, Chris Allton, and Jonas Glesaaen  for valuable helps, and thank Chris Allton for the discussion about the extrapolation method.
	We modify
	the adapted version of  OpenQCD code~\cite{openqcd} to carry out this simulation and we use the computer program~\cite{Hansen:code} to
	calculate the spectral density of correlation function.
	The adaptation of  OpenQCD code is  publicly available~\cite{fastsum1}. 
The simulations are carried out on $N_f=2+1$ Generation2 (Gen2)
FASTSUM ensembles~\cite{Aarts:2020vyb} of which the ensembles at the lowest temperature are provided by the HadSpec
collaboration~\cite{Edwards:2008ja,HadronSpectrum:2008xlg}.
This work
is supported by the National Natural Science Foundation of
China (NSFC) under Grant No. 11347029.
This work is done  at the high performance computing platform of Jiangsu University.

\end{acknowledgments}

\end{document}